\documentclass[twocolumn]{aastex701}
\usepackage{amsmath,amssymb}
\usepackage{graphicx}
\usepackage{dcolumn}
\usepackage{bm}
\usepackage{xcolor}
\usepackage{multirow}
\usepackage{booktabs}
\usepackage{longtable}
\usepackage{subcaption}
\usepackage{float}

\begin{document}

\title{Hierarchical Bayesian Inference on the intrinsic event rate of Type I gamma-ray bursts from the {\it Fermi}/GBM catalogue}

\correspondingauthor{Shu-Xu Yi}
\email[show]{sxyi@ihep.ac.cn}

\author[orcid=0009-0000-1101-3470,gname=Tian-Yong, sname=Cao]{Tian-Yong Cao} 
\affiliation{State Key Laboratory of Particle Astrophysics, Institute of High Energy Physics, Chinese Academy of Sciences, Beijing 100049, China}
\affiliation{University of Chinese Academy of Sciences, Chinese Academy of Sciences, Beijing 100049, People’s Republic of China}
\email{}

\author[orcid=0000-0001-7599-0174, gname=Shu-Xu, sname=Yi]{Shu-Xu Yi} 
\affiliation{State Key Laboratory of Particle Astrophysics, Institute of High Energy Physics, Chinese Academy of Sciences, Beijing 100049, China}
\affiliation{University of Chinese Academy of Sciences, Chinese Academy of Sciences, Beijing 100049, People’s Republic of China}
\email{}

\author[orcid=0000-0002-1905-1727, gname=Yun-Fei, sname=Du]{Yun-Fei Du}
\affiliation{State Key Laboratory of Particle Astrophysics, Institute of High Energy Physics, Chinese Academy of Sciences, Beijing 100049, China}
\affiliation{University of Chinese Academy of Sciences, Chinese Academy of Sciences, Beijing 100049, People’s Republic of China}
\affiliation{Beijing Academy Science and Technology Branch, No. 38 Huixinli, Chaoyang District, Beijing 100029, China}
\email{}

\author[orcid=0000-0002-8442-9458, gname=Emre, sname=S. Yorgancioglu]{Emre S. Yorgancioglu}
\affiliation{State Key Laboratory of Particle Astrophysics, Institute of High Energy Physics, Chinese Academy of Sciences, Beijing 100049, China}
\affiliation{University of Chinese Academy of Sciences, Chinese Academy of Sciences, Beijing 100049, People’s Republic of China}
\email{}

\author[orcid=0000-0002-4771-7653, gname=Shao-Lin, sname=Xiong]{Shao-Lin Xiong}
\affiliation{State Key Laboratory of Particle Astrophysics, Institute of High Energy Physics, Chinese Academy of Sciences, Beijing 100049, China}
\affiliation{University of Chinese Academy of Sciences, Chinese Academy of Sciences, Beijing 100049, People’s Republic of China}
\email{}


\begin{abstract}

\noindent The intrinsic event-rate of Type I gamma-ray bursts (GRBs) as a function of redshift, 
provides a probe for binary neutron star (BNS) mergers population. However, the above-mentioned inference is complicated by instrumental selection effects, the lack of redshift measurements for the majority of GRBs, and uncertainties in jet geometry. In this work, we applied a hierarchical Bayesian inference framework to the {\it Fermi}/GBM Type I GRB catalogue to reconstruct the underlying event-rate population. This framework accounts for the selection effects and marginalizes over the limited-information posteriors of individual GRB redshifts and jet structures. The population model uses parameterized luminosity function, distribution of jet structure parameters, the Madau-Dickinson star-formation rate and a delay-time distribution of BNS mergers. The posterior distribution of the hyperparameters in the model are to be inferred, using a GPU-accelerated MCMC algorithm. The intrinsic event rate density as function of redshift is constrained up to $z=5$, which a local rate density of \(R_0 = 22.6^{+15.0}_{-10.6}\,{\rm Gpc^{-3}\,yr^{-1}},\), which is consistent with the latest gravitational-wave measurements.

\end{abstract}



\section{Introduction}

Binary neutron star (BNS) mergers are widely regarded as one of the most important astrophysical sites of rapid neutron-capture (r-process) nucleosynthesis in the Universe \citep{lattimer1974black,eichler1989nucleosynthesis,rosswog1998mass,drout2017light}. As a result, reconstructing the cosmic merger rate history of BNS systems is crucial for understanding the origin and chemical evolution of heavy elements. It also provide crucial clues to physics of stellar evolution \citep{dominik2012double,belczynski2016first,nelemans2001gravitational,nelemans2009galactic}. Furthermore, an accurate characterization of the BNS merger rate across cosmic time plays a key role in forecasting detection rates and shaping observational strategies for future gravitational-wave (GW) and multi-messenger astronomy \citep{abadie2010predictions,morinaga2019statistical,radice2020dynamics}.

Direct constraints on the BNS merger rate have been provided by the observations of the LIGO–Virgo–KAGRA (LVK) Collaboration. Successive GW transient catalogs, including GWTC-2, GWTC-3, GWTC-4.0 and the latest GWTC-5.0, report merger rates spanning between \(320_{-240}^{+490}\) \citep{abbott2021population}, \(10\text{–}1700\) \citep{abbott2023population}, \(7.6\text{–}250\,\mathrm{Gpc^{-3}\,yr^{-1}}\) \citep{abac2025gwtc}, and \(5.1\text{–}154.7\,\mathrm{Gpc^{-3}\,yr^{-1}}\) \citep{ligo2026gwtc}. These estimations are limited to the local Universe and suffer from large uncertainties due to small event number. 

Long before the advent of GW astronomy, a class of gamma-ray bursts (GRBs), commonly referred to as Type I GRBs (or Short GRBs), had been hypothesized to originate from compact binary mergers, particularly BNS systems \citep[e.g.][]{berger2014short}. This association was firmly established by the joint detection of GW170817 and its electromagnetic counterpart GRB170817A \citep{abbott2017gw170817}. Nevertheless, recent observational studies have suggested that the traditional classification of GRBs based solely on duration (i.e., short vs. long) may not reliably reflect their physical progenitors  \citep{yi2025long,rastinejad2022kilonova,levan2024heavy,yang2022boosting,sun2025magnetar,wang2025subclass}. Indeed, there is subclass of merger-origin GRB which lasts much longer than 2 seconds, dubbed Type IL GRB \citep{wang2025subclass}. On the other hand, some short burst may be produced by collapsar, rather than compact binary merger.

Given the established link between Type I GRBs and BNS mergers, along with the much larger sample size and greater redshift reach of GRB observations, these observations have been widely used as an independent and complementary probe to infer the merger rate of BNSs \citep{wanderman2015rate,fong2015decade,rouco2023jet,coward2012swift,dominik2013double}. By modeling the luminosity function, redshift distribution, and selection effects of GRBs, previous studies have attempted to reconstruct the cosmic evolution of the BNS merger rate, providing valuable insights that extend beyond the limitations of current GW observations. 

Despite these efforts, constraining the intrinsic Type I GRB rate as a function of redshift remains highly challenging. Previous studies are subject to several systematic uncertainties. First, instrumental selection effects—arising from detector sensitivity thresholds, energy band limitations, and complex trigger algorithms—can significantly bias the observed GRB sample, particularly against low-luminosity or high-redshift events. Second, the redshift completeness of GRB samples is still limited, as only a fraction of bursts have reliable redshift measurements due to observational constraints such as afterglow brightness and follow-up capabilities \citep{jakobsson2006mean,fynbo2009low,berger2014short}. This incompleteness can introduce nontrivial biases in reconstructing the underlying redshift distribution. Third, the jet opening angle of GRBs, which is required to convert the observed event rate into the true all-sky rate, remains poorly constrained and may exhibit intrinsic diversity across the population \citep{fong2015decade,berger2014short,ghirlanda2016short}. Additional uncertainties, such as the possible evolution of GRB luminosity functions and the ambiguity in progenitor classification, further complicate the interpretation. As a result, it is difficult to directly infer an intrinsic Type I GRB rate that faithfully traces the cosmic merger rate history of binary neutron stars \citep{wanderman2015rate,d2015short}.

Recently, \citealt{salafia2023short} (S23, hereafter) applied a hierarchical Bayesian framework to short GRBs to address selection effects and incomplete observational information. Their analysis, however, relies on several simplifying assumptions, including a universal jet structure shared by all sources, a simplified step-function approximation for the detection threshold, and a merger-rate model that does not account for the time delay between binary formation and merger. 

In this work, we develop a GPU-accelerated hierarchical Bayesian framework and relax these assumptions by adopting a more flexible modeling strategy. We allow for intrinsic diversity in the jet structure by assuming that the half-opening angles of structured jets vary across sources and follow a Gaussian distribution. We model the BNS merger rate as a redshift-dependent broken power law convolved with a time-delay distribution to account for the lag between binary formation and merger. To mitigate the impact of incomplete redshift measurements, we incorporate empirical correlations to statistically infer missing redshift information. For the detection probability, instead of adopting a simplified step-function approximation, we employ the fitted detection efficiency model from \cite{howell2025apparent}, which provides a more realistic description of the instrument response. Furthermore, when mapping theoretical GRB models to their observable counterparts in the detector, we explicitly account for stochastic spikiness calibrated from real observational data, thereby improving the fidelity of the forward modeling.

The remainder of this paper is organized as follows: in Section 2, we describe the hierarchical Bayesian framework, data treatment, and modeling assumptions; in Section 3, we present the inferred population properties; and in Section 4, we discuss the implications and limitations of our results.

\section{Methodology}
\begin{figure}
    \centering
    \includegraphics[width=\linewidth]{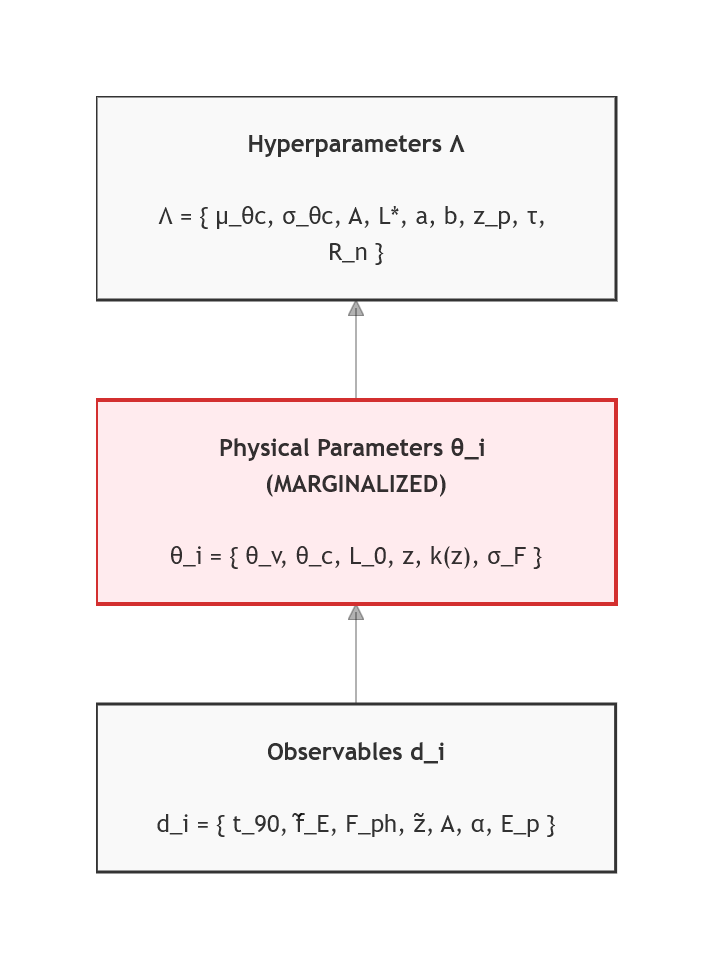}
    \caption{The framework of the Hierarchical Bayesian framework in this work.}
    \label{fig:framework}
\end{figure}

Our Hierarchical Bayesian framework is illustrated in figure \ref{fig:framework}. In the following subsections, we will introduce in detail the equations, and parameters in each layers. 

\subsection{Hierarchical Bayesian}
\label{sect:HB}

When a GRB is detected, the instrument provides a set of observables \(\Vec{d}\), which can be described by a set of physical parameters \(\Vec{\theta}\). Within the standard Bayesian framework, the posterior distribution of the parameters is given by:
\begin{equation}
p(\Vec{\theta}|\Vec{d}) = \frac{\mathcal{L}(\Vec{\theta}|\Vec{d})\,\pi(\Vec{\theta})}{p(\Vec{d})}\,,
\end{equation}
where \(\mathcal{L}(\Vec{\theta}|\Vec{d})\) is the likelihood, \(\pi(\Vec{\theta})\) is the prior, and \(p(\Vec{d})\) is the evidence. For a fixed model, the evidence serves as a normalization constant and is often omitted.

In population studies, each event \(i\) is characterized by its own parameters \(\Vec{\theta}_i\), which are assumed to be drawn from an underlying population distribution \(p_{\rm pop}(\Vec{\theta}|\Vec{\Lambda})\), governed by hyperparameters \(\Vec{\Lambda}\). Treating this population distribution as the prior on \(\Vec{\theta}_i\), and assigning a prior \(\pi(\Vec{\Lambda})\) to the hyperparameters, the joint posterior becomes
\begin{equation}
p(\Vec{\theta}_i, \Vec{\Lambda}|\Vec{d}_i) \propto \mathcal{L}(\Vec{\theta}_i|\Vec{d}_i)\, p_{\rm pop}(\Vec{\theta}_i|\Vec{\Lambda})\pi(\Vec{\Lambda})\,.
\end{equation}

\begin{figure*}
    \centering
    \includegraphics[width=\linewidth]{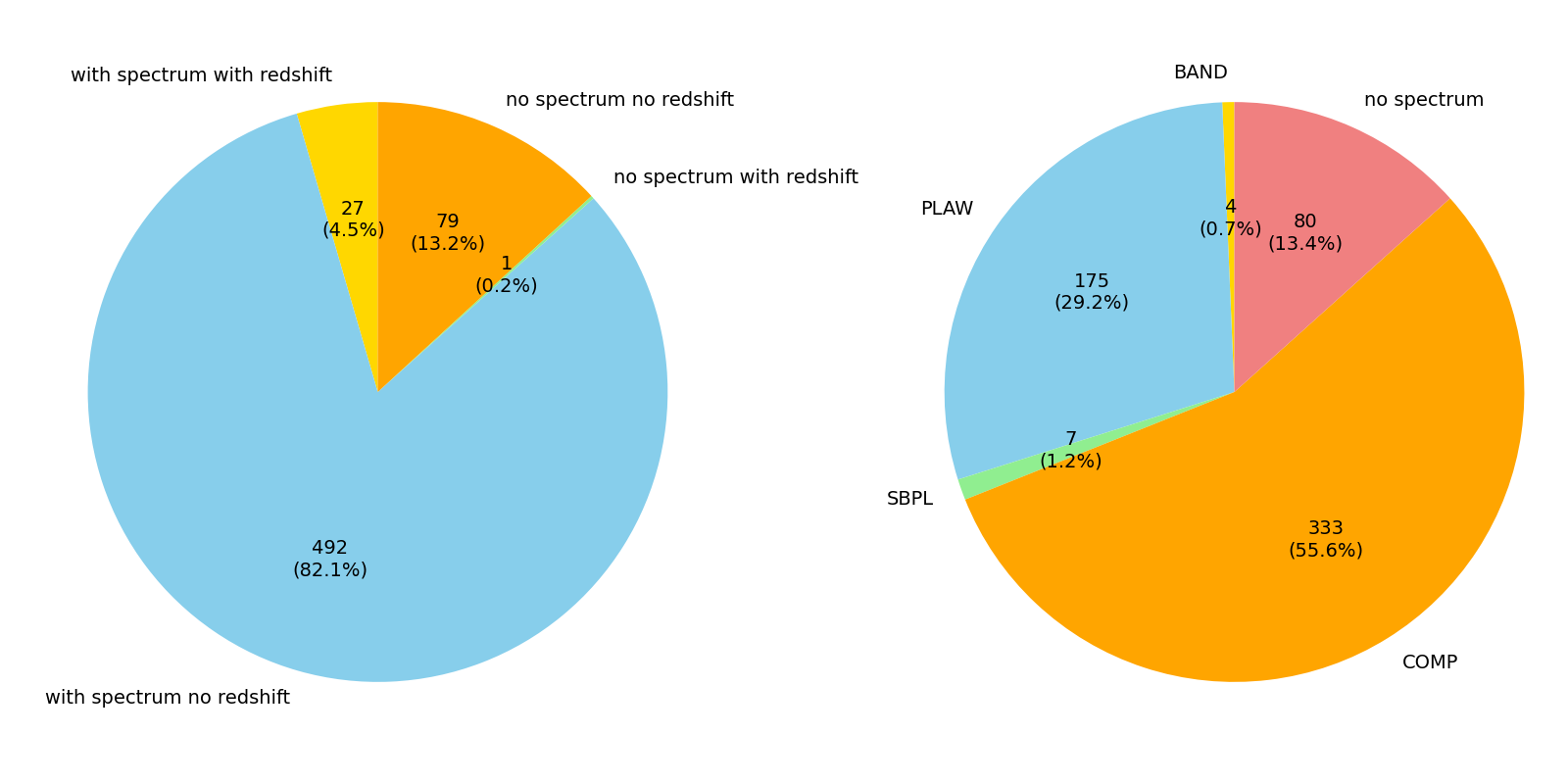}
    \caption{Fractions of different source types in the {\it Fermi}/GBM catalog}
    \label{fig:pie}
\end{figure*}

In practice, we are often interested in the hyperparameters \(\Vec{\Lambda}\), which can be obtained by marginalizing over the individual event parameters. For a set of \(N_{\rm obs}\) observed events, the posterior can be written as a Poisson process \citep{vitale2022inferring}:
\begin{equation}
\begin{aligned}
p(\Vec{\Lambda}|\Vec{d}) \propto & \pi(\Vec{\Lambda})N_{\rm tot}(\Vec{\Lambda})^{N_{\rm obs}}e^{-N_{\rm tot}(\Vec{\Lambda})}\\ 
&\cdot\prod_{i=1}^{N_{\rm obs}}\int \mathrm{d}\Vec{\theta_i} \mathcal{L}(\Vec{\theta}_i|\Vec{d}_i)\, p_{\rm pop}(\Vec{\theta}_i|\Vec{\Lambda})\\
= &\pi(\Vec{\Lambda})N_{\rm tot}(\Vec{\Lambda})^{N_{\rm obs}}e^{-N_{\rm tot}(\Vec{\Lambda})} \prod_{i=1}^{N_{\rm obs}}\mathcal{L}(\Vec{\Lambda}|\Vec{d}_i)\,,
\end{aligned}
\end{equation}
where \(N_{\rm tot}(\Vec{\Lambda})\) denotes the total number of sources in the Universe. The prefactor arises from the Poisson probability of observing \(N_{\rm obs}\) events.

However, detectors can only observe a fraction of all sources due to selection effects: intrinsically brighter or closer events are more likely to be detected. To account for this bias, the likelihood must be corrected as \citep{messick2017analysis}:
\begin{equation}
p(\Vec{\Lambda}|\Vec{d}) \propto
\pi(\Vec{\Lambda})N_{\rm exp}(\Vec{\Lambda})^{N_{\rm obs}}
e^{-N_{\rm exp}(\Vec{\Lambda})}
\prod_{i=1}^{N_{\rm obs}}
p(\Vec{d}_i|{\rm Det}, \Vec{\Lambda})\,,
\end{equation}
where \(N_{\rm exp}(\Vec{\Lambda})\) is the expected number of detected events:
\begin{equation}
N_{\rm exp}(\Vec{\Lambda}) = 0.6N_{\rm tot}(\Vec{\Lambda})\xi(\Vec{\Lambda})\,.
\end{equation}
Here, \(\xi(\Vec{\Lambda})\) is the detection efficiency, and the factor \(0.6\) accounts for the average sky coverage of {\it Fermi}/GBM.

The likelihood conditioned on detection can be written as \citep{loredo2004accounting}:
\begin{equation}
p(\Vec{d}|{\rm Det}, \Vec{\Lambda}) =
\frac{p({\rm Det}|\Vec{d}, \Vec{\Lambda}) \mathcal{L}(\Vec{\Lambda}|\Vec{d})}
{p({\rm Det}|\Vec{\Lambda})}\,,
\label{like_1}
\end{equation}
where the denominator represents the probability that an event is detected under the population model,
\begin{equation}
p({\rm Det}|\Vec{\Lambda}) = \xi(\Vec{\Lambda})\,.
\end{equation}

In practice, detection is determined by a statistic constructed from the observables \(\Vec{d}\), such as the signal-to-noise ratio or false-alarm rate, which we denote as \(\Theta(\Vec{d})\). Given a detection threshold \(\Theta_0\), the detection probability can be approximated as a step function,
\begin{equation}
p({\rm Det}\perp\Vec{\theta}|\Vec{d}) =
\begin{cases}
1, & \Theta(\Vec{d}) > \Theta_0 \\
0, & \Theta(\Vec{d}) < \Theta_0
\end{cases}\,.
\end{equation}
In this case, the detectability of a source is independent of its physical parameters \(\Vec{\theta}\) and depends solely on its observables \(\Vec{d}\). We denote this independence by \(\perp \Vec{\theta}\). 

For a given observed event, \(\Vec{d}\) is fixed, and thus \(p({\rm Det}|\Vec{d}, \Vec{\Lambda})\) reduces to a constant \citep{essick2024ensuring}. In this case, Eq.~\ref{like_1} simplifies to:
\begin{equation}
p(\Vec{d}|{\rm Det}, \Vec{\Lambda})
\propto
\frac{\mathcal{L}(\Vec{\Lambda}|\Vec{d})}
{p({\rm Det}|\Vec{\Lambda})}\,.
\end{equation}

The numerator can be written by marginalizing over the latent parameters:
\begin{equation}
\mathcal{L}(\Vec{\Lambda}|\Vec{d}) = \int \mathrm{d}\Vec{\theta}
\mathcal{L}(\Vec{\theta}|\Vec{d})
p_{\rm pop}(\Vec{\theta}|\Vec{\Lambda})\,,
\end{equation}
while the denominator formally involves a double marginalization over both \(\Vec{\theta}\) and \(\Vec{d}\),
\begin{equation}
p({\rm Det}|\Vec{\Lambda}) = \iint \mathrm{d}\Vec{\theta}\mathrm{d}\Vec{d}
\mathcal{L}(\Vec{\theta}|\Vec{d})
p({\rm Det}|\Vec{d})
p_{\rm pop}(\Vec{\theta}|\Vec{\Lambda})\,.
\end{equation}
Here, \(\Vec{d}\) in the numerator denotes the actual observed data, whereas in the denominator it is a latent variable to be marginalized over. Evaluating this integral typically requires simulating a large number of mock observations \(\Vec{d}\) and computing the corresponding detection statistic \(\Theta(\Vec{d})\), which can be computationally expensive.

To alleviate this cost, one can approximate the detection probability as a function of the intrinsic parameters. For GRBs, \cite{howell2025apparent} modeled the detection probability as a function of the peak photon flux \(F_{\rm ph}\) (measured in \(\mathrm{photon}\cdot\mathrm{s}^{-1}\mathrm{cm}^{-2}\) on a 64 ms timescale). Since \(F_{\rm ph}\) can be predicted from \(\Vec{\theta}\), we approximate:
\begin{equation}
p({\rm Det}\perp\Vec{d}|\Vec{\theta}) = p_{\rm det}(F_{\rm ph})\,.
\end{equation}
In this case, whether a source is detected depends only on its physical parameters \(\Vec{\theta}\) and is independent of the observables \(\Vec{d}\). For a more detailed discussion on this point, see \cite{essick2024ensuring}.

Under this approximation, the detection probability no longer depends explicitly on \(\Vec{d}\), and the denominator reduces to a single integral over the population distribution,
\begin{equation}
p({\rm Det}|\Vec{\Lambda}) = \int \mathrm{d}\Vec{\theta}
p_{\rm det}(F_{\rm ph})
p_{\rm pop}(\Vec{\theta}|\Vec{\Lambda})\,.
\end{equation}

Similarly, the numerator of Eq.~\ref{like_1} becomes:
\begin{equation}
\begin{aligned}
p({\rm Det}|\Vec{d},\Vec{\Lambda})\mathcal{L}(\Vec{\Lambda}|\Vec{d})
&= \frac{p({\rm Det}, \Vec{d}|\Vec{\Lambda})}{p(\Vec{d}|\Vec{\Lambda})}\mathcal{L}(\Vec{\Lambda}|\Vec{d}) \\
&= \int \mathrm{d}\Vec{\theta}
p({\rm Det}, \Vec{d}|\Vec{\theta})
p_{\rm pop}(\Vec{\theta}|\Vec{\Lambda}) \\
&= \int \mathrm{d}\Vec{\theta}
p({\rm Det}|\Vec{\theta})
\mathcal{L}(\Vec{\theta}|\Vec{d})
p_{\rm pop}(\Vec{\theta}|\Vec{\Lambda})\,.
\end{aligned}
\end{equation}

Therefore, the likelihood conditioned on detection can be written as
\begin{equation}
p(\Vec{d}|{\rm Det}, \Vec{\Lambda}) =
\frac{\int \mathrm{d}\Vec{\theta}
p_{\rm det}(F_{\rm ph})
\mathcal{L}(\Vec{\theta}|\Vec{d})
p_{\rm pop}(\Vec{\theta}|\Vec{\Lambda})}
{\int \mathrm{d}\Vec{\theta}
p_{\rm det}(F_{\rm ph})
p_{\rm pop}(\Vec{\theta}|\Vec{\Lambda})}\,,
\label{like_2}
\end{equation}
which is the form adopted in this work.

\subsection{Data Selection and Preparation}

In this work, we analyze the {\it Fermi}/GBM GRB catalog \citep{von2020fourth,gruber2014fermi,von2014second,bhat2016third}, using data available up to August 18, 2023\footnote{The online catalog is available at: \url{https://heasarc.gsfc.nasa.gov/w3browse/fermi/fermigbrst.html}}. The catalog provides several observable properties of GRBs. In this study, we utilize the duration over which 90\% of the burst fluence is accumulated, \(\Tilde{t}_{90}\) (in seconds), the fluence integrated over the burst duration in the \(50\text{--}300\,\rm keV\) energy band, \(\Tilde{f}_{\rm E}\) (in units of \(\rm erg\,cm^{-2}\)), and the peak photon flux measured on a 64 ms timescale in the same energy band, \(\Tilde{F}_{\rm ph}\) (in units of \(\mathrm{photon}\,\mathrm{s}^{-1}\,\mathrm{cm}^{-2}\)). To distinguish these measured quantities from those inferred from physical parameters \(\Vec{\theta}\), we denote all observables with a tilde.

We define the observed energy flux averaged over the burst duration as
\begin{equation}
\Tilde{F}_{\rm E} = \frac{\Tilde{f}_{\rm E}}{\Tilde{t}_{90}}\,,
\end{equation}
where \(\Tilde{F}_{\rm E}\) has units of \(\rm erg\,cm^{-2}\,s^{-1}\).

The quantities \(\Tilde{F}_{\rm E}\) and \(\Tilde{F}_{\rm ph}\) differ not only
in their units, but also in their definitions. The former represents the duration-averaged energy flux inferred from the fluence and \(t_{90}\), while
the latter corresponds to the peak photon flux measured with a time resolution of 64 ms, thereby preserving more rapid temporal variability. In the following,
we refer to \(\Tilde{F}_{\rm E}\) as the duration-averaged energy flux, and \(\Tilde{F}_{\rm ph}\) as the 64 ms-bin peak photon flux.

\begin{table}[]
\centering
\renewcommand{\arraystretch}{1.5}
\caption{The table of GRBs with redshift}
\setlength{\tabcolsep}{36pt}
\begin{tabular}{cc}
\hline
GBM\_name    & \(\Tilde{z}\)        \\ \hline
GRB241209233 & 1.49     \\
GRB201221963 & 1.046    \\
GRB160624477 & 0.483    \\
GRB150101641 & 0.093    \\
GRB131004904 & 0.717    \\
GRB100816026 & 0.8034   \\
GRB090927422 & 1.37     \\
GRB090510016 & 0.903    \\
GRB080905499 & 0.1218   \\
GRB100117879 & 0.92     \\
GRB111117510 & 2.211    \\
GRB160821937 & 0.16     \\
GRB200826187 & 0.7481   \\
GRB081024891 & 3.05     \\
GRB081024245 & 3.05     \\
GRB100206563 & 0.4068   \\
GRB100625773 & 0.452    \\
GRB140619490 & 2.67     \\
GRB150101270 & 0.093    \\
GRB170817529 & 0.009783 \\
GRB201221591 & 1.045    \\
GRB130515056 & 0.8      \\
GRB170127634 & 2.28     \\
GRB180727594 & 1.95     \\
GRB191031891 & 1.93     \\
GRB200219317 & 0.48     \\
GRB200411187 & 0.82     \\
GRB210323918 & 0.733    \\ \hline
\end{tabular}
\label{tab:redshift}
\end{table}

The catalog contains a total of 3598 GRB events. Following \cite{howell2025apparent}, we adopt \(t_{90} = 2.1\,\rm s\) as the classification threshold in order to ensure consistency with the detection probability curve used in our analysis. We identify 604 short GRBs, all of which we assume to be Type I GRBs. Based on the detection probability curve derived in \cite{howell2025apparent}, whose minimum corresponds to \(F_{\rm ph} = 0.5217\,\mathrm{photon}\,\mathrm{s}^{-1}\,\mathrm{cm}^{-2}\), we conservatively assume that GRBs with peak photon flux below this threshold have zero detection probability.  Consequently, we exclude 5 events with \(\Tilde{F}_{\rm ph} < 0.5217\,\mathrm{photon}\,\mathrm{s}^{-1}\,\mathrm{cm}^{-2}\), leaving a final sample of 599 GRBs for the hierarchical Bayesian analysis.


\begin{figure}
    \centering
    \includegraphics[width=\linewidth]{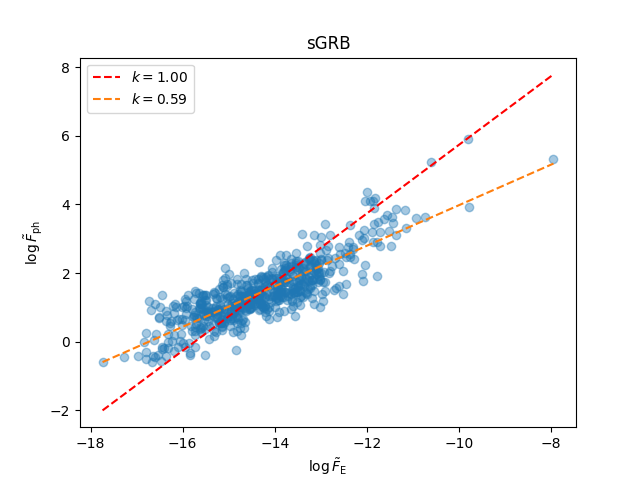}
    \caption{Relation between peak photon flux in 64 ms and duration-averaged energy flux in the Fermi catalog: The red curves represent the fitting results obtained with \(k_F = 1\) fixed, and the orange curves represent the fitting results with \(k_F\) treated as a free parameter.}
    \label{fig:flux_fit}
\end{figure}

For each GRB, {\it Fermi}/GBM performs spectral fitting using four candidate models: Power Law (PLAW), Comptonized (COMP), Band (BAND), and Smoothly Broken Power Law (SBPL), with:

\noindent\(\bullet\) PLAW:
\begin{equation}
N(E|A,\alpha)=A\big(\frac{E}{100\,{\rm keV}}\big)^\alpha\,;
\end{equation}
\noindent\(\bullet\) COMP:
\begin{equation}
N(E|A,\alpha,E_{\rm p})=A\big(\frac{E}{100\,{\rm keV}}\big)^\alpha \exp\big(-\frac{(2+\alpha)E}{E_{\rm p}}\big)\,;
\end{equation}
\noindent\(\bullet\) BAND:
\begin{equation}
\begin{aligned}
&N(E|A,E_{\rm p},\alpha,\beta) \\
&= 
\begin{cases}
&A \big(\frac{E}{100\,\mathrm{keV}}\big)^{\alpha} \exp\big(-\frac{E}{E_{\rm c}}\big), E < (\alpha-\beta) E_{\rm c}\\ 
&A \big(\frac{(\alpha-\beta)E_{\rm c}}{100\,\mathrm{keV}}\big)^{\alpha-\beta} e^{(\beta-\alpha)}\big(\frac{E}{100\,\mathrm{keV}}\big)^\beta, E \ge (\alpha-\beta) E_{\rm c}\\
&E_{\rm c} = (\alpha+2) E_{\rm p}
\end{cases}\,;
\end{aligned}
\end{equation}
\noindent\(\bullet\) SBPL:
\begin{equation}
\begin{aligned}
&N(E|A,\alpha,E_{\rm b},\beta)\\
&= A\big(\frac{E}{100\,{\rm keV}}\big)^{\frac{\alpha+\beta}{2}}\Big[\frac{\cosh{\log(E/E_{\rm b})\over0.3}}{\cosh{\log(100\,{\rm keV}/E_{\rm b})\over0.3}}\Big]^{\frac{\beta-\alpha}{2}}\,.
\end{aligned}
\end{equation}
The best-fitting model is selected based on the likelihood-based Cash statistic (CSTAT). These criteria follow the Fermi/GBM spectral catalog prescription \citep{goldstein2012fermi}, as summarized in the HEASARC documentation.\footnote{\url{https://heasarc.gsfc.nasa.gov/w3browse/fermi/fermigbrst.html}} Specifically, the COMP model is preferred over PLAW if the CSTAT decreases by at least 8.58. The BAND or SBPL models are preferred over COMP if the CSTAT decreases by at least 11.83, with the model yielding the lower CSTAT between BAND and SBPL being selected. Additionally, a model is considered acceptable only if its parameters are well constrained: the low-energy spectral index must be determined to within 0.4 (68\% confidence level), the high-energy index within 1.0, and all other parameters within 40\% of their best-fit values. Based on these criteria, 333 GRBs are best fitted by the COMP model, 175 by PLAW, 7 by SBPL, and 4 by BAND, while 80 GRBs lack reliable spectral fits. Detailed information for each source can be found in Appendix~\ref{app_5}.

Since the {\it Fermi}/GBM catalog does not provide redshift measurements, we compile redshift information from the {\it Swift}/BAT catalog \citep{barthelmy2005burst} and the literature \cite{lan2023grb,salafia2023short}. By cross-matching burst times and sky localizations, we identify 28 GRBs in the {\it Fermi}/GBM sample with associated redshift measurements, as listed in Table~\ref{tab:redshift}.

Figure~\ref{fig:pie} summarizes the distribution of GRBs across different spectral models and the availability of redshift information.

\subsection{GRB model}
\label{GRB model}

In this work, we assume that the GRB jet follows a Gaussian structured jet model, and allow the parameters of the jet structure to vary from burst to burst. In the observer frame, the luminosity of the jet as a function of the viewing angle \(\theta_v\) can be written as \citep{guo2020luminosity}:
\begin{equation}
L(\theta_v,\theta_c,L_0) = L_0\exp\left(-\frac{\theta_v^2}{\theta_c^2}\right)\,,
\label{eq:L}
\end{equation}
where \(\theta_c\) is the characteristic angle (core angle) of the jet, and \(L_0\) denotes the isotropic-equivalent luminosity averaged over duration when the jet is observed along its axis. After applying the \(k\)-correction \(k(z)\), energy flux reaching the detector can be written as:
\begin{equation}
F_{\rm E} = \frac{L(\theta_v,\theta_c,L_0)}{4\pi D_L^2(z)k(z)}\,.
\end{equation}
It should be noted that the quantity \(F_{\rm E}\) obtained here is duration-averaged energy flux. However, when calculating the detection probability, we need to use 64ms-bin photon flux \(F_{\rm ph}\). Therefore, it is necessary to establish a conversion relation between these two quantities.

\begin{table}[]
\centering
\renewcommand{\arraystretch}{1.5}
\caption{Fitting relations between peak photon flux and peak energy flux for different types of Type I GRBs: \(\mathcal{N}[\mu, \sigma]\) denotes a normal distribution with mean \(\mu\) and variance \(\sigma^2\).}
\setlength{\tabcolsep}{26pt}
\begin{tabular}{ccc}
\hline
 \(k_F\) & \(b_F\) & \(\sigma_F\)                          \\ \hline
 1\,(fixed)       & 16.318  & \(\mathcal{N}[0, 0.70]\) \\
 0.59    & 10.228    & \(\mathcal{N}[0, 0.47]\) \\ \hline
\end{tabular}
\label{tab:flux_fit}
\end{table}

The {\it Fermi}/GRB catalog \citep{von2020fourth,gruber2014fermi,von2014second,bhat2016third} provides both of these observables. We compare \(F_{\rm E}\) and \(F_{\rm ph}\) for all GRBs in the catalog using a scatter plot, as shown in the Figure \ref{fig:flux_fit}. It can be seen that in logarithmic space the two quantities approximately follow a linear relation, with some intrinsic scatter:
\begin{equation}
\ln F_{\rm ph}(F_{\rm E}, \sigma_F) = k_F \ln F_{\rm E} + b_F + \sigma_F\,.
\end{equation}
The difference between the two corresponds to an average photon energy \(C_N\). Thus, we assume that \(F_{\rm ph}\) is linearly proportional to \(F_{\rm E}\), i.e., we fix \(k_F = 1\). Under this assumption, the fitted value of \(b_F\) and the distribution of \(\sigma_F\) are listed in the Table \ref{tab:flux_fit}. In addition, we also allow \(k_F\) to be a free parameter and perform the fit again, and the results are also presented in the Table \ref{tab:flux_fit}.

In the above relation, \(\sigma_F\) represents a source-by-source scatter parameter. Since this work does not focus on the specific value of \(\sigma_F\) for each event, we marginalize over it when calculating the detection probability, i.e.,
\begin{equation}
p_{\rm det}(F_{\rm ph}) =
\int \mathrm{d}\sigma_F
p_{\rm det}\big(F_{\rm ph}(F_{\rm E}, \sigma_F)\big)
p(\sigma_F)\,,
\label{eq:pdet}
\end{equation}
here \(p(\sigma_F)\) denotes the empirical distribution of \(\sigma_F\) obtained from fitting the Fermi GRB catalog.

In summary, within the GRB model adopted in this work, the physical parameters of GRBs include: the viewing angle \(\theta_v\), the jet half-opening angle \(\theta_c\), the isotropic-equivalent luminosity along the jet axis \(L_0\), the redshift \(z\), and the \(k\)-correction factor \(k(z)\), i.e., \(\Vec{\theta} = \{\theta_v,\theta_c,L_0,z,k(z)\}\)\footnote{It is worth noting that the complete parameter vector \(\Vec{\theta}\) should include \(\sigma_F\). Since this parameter is not of primary interest, we marginalize over it in the discussion in Section~\ref{GRB model}. Therefore, in the following discussion, we no longer include it as part of \(\Vec{\theta}\).}. 

For the observable \(\vec{d}\), sources that have neither spectral information nor redshift measurements provide only \(\tilde{F}_{\rm E}\). If spectral information is available, \(\vec{d}\) additionally includes the k-correction factor \(\tilde{k}(z)\) (note that \(\tilde{k}(z)\) is not a single value, but a redshift-dependent function determined by the spectral information). When an independent redshift measurement is available, \(\vec{d}\) further includes the measured value \(\tilde{z}\), i.e., \(\Vec{d} = \{\Tilde{F}_{\rm E}, \Tilde{z}, \Tilde{k}(z)\}\).

\subsection{Likelihood}

In this subsection, we briefly introduce the likelihood functions constructed from observations.

\(\bullet\) Duration-averaged energy flux \(F_{\rm E}\):
The measured value \(\Tilde{F}_{\rm E}\) and its uncertainty \(\Delta \Tilde{F}_{\rm E}\) can be directly obtained from the {\it Fermi}/GBM catalog. We assume a log-normal likelihood function,
\begin{equation}
\mathcal{L}\big(F_{\rm E}(\Vec{\theta}) \mid \Tilde{F}_{\rm E}\big) =
\mathcal{N}\left[\log \Tilde{F}_{\rm E}\,, \Delta \Tilde{F}_{\rm E}/\Tilde{F}_{\rm E}\right]\big(\log F_{\rm E}(\Vec{\theta})\big)\,,
\end{equation}
where \(\mathcal{N}[\mu, \sigma](x)\) denotes a normal distribution of \(x\) with mean \(\mu\) and standard deviation \(\sigma\).

\(\bullet\) \(k\)-correction factor \(k(z)\):
For events with spectral information, the \(k\)-correction factor is computed as:
\begin{equation}
\Tilde{k}(z)=
\frac{\int_{E_1}^{E_2} N(E \mid \Tilde{\Vec{\lambda}})\,\mathrm{d}E}
{\int_{(1+z)E_1}^{(1+z)E_2} N(E \mid \Tilde{\Vec{\lambda}})\,\mathrm{d}E}\,,
\end{equation}
where \(N(E \mid \Tilde{\Vec{\lambda}})\) is the spectral model with measured parameters \(\Tilde{\Vec{\lambda}}\) (see Appendix~\ref{app_5}). Throughout this work, we adopt a uniform energy band of \(50\text{--}300\,\mathrm{keV}\), i.e., \(E_1 = 50\,\mathrm{keV}\) and \(E_2 = 300\,\mathrm{keV}\).

After computing \(\Tilde{k}(z)\), we neglect its uncertainty and model its likelihood as a Dirac delta function,
\begin{equation}
\mathcal{L}\big(k(z) \mid \Tilde{k}_z\big) = \delta\big(k(z) - \Tilde{k}_z\big)\,.
\end{equation}

\(\bullet\) Redshift \(z\):
For events with measured redshift, we assume that the observational uncertainty is negligible. The likelihood is therefore:
\begin{equation}
\mathcal{L}(z \mid \Tilde{z}) = \delta(z - \Tilde{z})\,.
\end{equation}

As shown in Figure~\ref{fig:pie}, only about \(5\%\) of the total sample has measured redshifts. Consequently, the statistical constraints from these events alone are limited. To improve the inference of the redshift distribution, we further incorporate empirical relationships (EMRs), which is discussed in Appendix \ref{emrs}. 

\(\bullet\) Half-opening angle \(\theta_c\) and viewing angle \(\theta_v\): For jets with observationally constrained half-opening angles, we model the constraint with a Dirac delta function,
\begin{equation}
    \mathcal{L}(\theta_c|\tilde{\theta}_c)=
    \delta(\theta_c-\tilde{\theta}_c)\,.
\end{equation}
The same treatment is adopted for the viewing angle,
\begin{equation}
    \mathcal{L}(\theta_v|\tilde{\theta}_v)=
    \delta(\theta_v-\tilde{\theta}_v)\,.
\end{equation}

Currently, we include only one event, GRB 170817529. Its jet core half-opening angle is constrained to be of order \(\theta_c \simeq 0.08~{\rm rad}\), while the viewing angle is \(\theta_v \simeq 0.35~{\rm rad}\), based on the VLBI superluminal-motion measurement and afterglow modeling \citep{mooley2018superluminal}.

Finally, the full likelihood depends on the available observational information:

\noindent \(\bullet\) Without spectral data:
\begin{equation}
\mathcal{L}(\Vec{\theta} \mid \Vec{d}) =
\mathcal{L}\big(F_{\rm E}(\Vec{\theta}) \mid \Tilde{F}_{\rm E}\big)\,.
\end{equation}

\noindent \(\bullet\) With spectral data but without redshift:
\begin{equation}
\mathcal{L}(\Vec{\theta} \mid \Vec{d}) =
\mathcal{L}\big(F_{\rm E}(\Vec{\theta}) \mid \Tilde{F}_{\rm E}\big)\,
\mathcal{L}\big(k(z) \mid \Tilde{k}_z\big)\,
\mathcal{L}\big(z \mid \Tilde{E}_{\rm iso}, \Tilde{E}_{\rm p}, \Tilde{t}_{90}\big)\,.
\end{equation}

\noindent \(\bullet\) With redshift but without spectral data:
\begin{equation}
\mathcal{L}(\Vec{\theta} \mid \Vec{d}) =
\mathcal{L}\big(F_{\rm E}(\Vec{\theta}) \mid \Tilde{F}_{\rm E}\big)\,
\mathcal{L}(z \mid \Tilde{z})\,.
\end{equation}

\noindent \(\bullet\) With both spectral data and redshift:
\begin{equation}
\mathcal{L}(\Vec{\theta} \mid \Vec{d}) =
\mathcal{L}\big(F_{\rm E}(\Vec{\theta}) \mid \Tilde{F}_{\rm E}\big)\,
\mathcal{L}\big(k(z) \mid \Tilde{k}_z\big)\,
\mathcal{L}(z \mid \Tilde{z})\,.
\end{equation}

\noindent \(\bullet\) GRB170817529:
\begin{equation}
\mathcal{L}(\Vec{\theta} \mid \Vec{d}) =
\mathcal{L}\big(F_{\rm E}(\Vec{\theta}) \mid \Tilde{F}_{\rm E}\big)\,
\mathcal{L}\big(k(z) \mid \Tilde{k}_z\big)\,
\mathcal{L}(z \mid \Tilde{z})\mathcal{L}(\theta_c|\tilde{\theta}_c)\,.
\end{equation}

\subsection{Population model}

In this subsection, we briefly introduce the population distribution models of these parameters and their corresponding hyperparameters.

\(\bullet\) Viewing angle \((\theta_v)\): We assume that the GRB jet directions are isotropically distributed in space; therefore, the viewing angle follows:
\begin{equation}
p_{\rm pop}(\theta_v) = \sin{\theta_v}\,.
\end{equation}

\(\bullet\) Jet half-opening angle \((\theta_c)\): We assume that it follows a log-normal distribution with mean \(\mu_{\theta_c}\) and standard deviation \(\sigma_{\theta_c}\), i.e.,
\begin{equation}
p_{\rm pop}(\theta_c|\mu_{\theta_c},\sigma_{\theta_c}) = \mathcal{N}(\log\theta_c|\log\mu_{\theta_c},\sigma_{\theta_c})\,.
\label{eq:theta_c_pop}
\end{equation}

\(\bullet\) On-axis luminosity \((L_0)\): We assume that it follows a power-law distribution with index \(-A\), with an exponential cutoff introduced at the low-luminosity end \(L_*\). Its form is \citep{wanderman2015rate,tan2020jet}:
\begin{equation}
p_{\rm pop}(L_0|A,L_*)=\frac{A}{L_*\Gamma(1-1/A)}
\exp\left[-\left(\frac{L_*}{L_0}\right)^A\right]
\left(\frac{L_0}{L_*}\right)^{-A}\,,
\end{equation}
where \(\Gamma(x)\) denotes the Gamma function.

\(\bullet\) Redshift \(z\): For Type I GRBs, the burst time is offset from the time of star-formation rate (SFR) by a delay:
\begin{equation}
R(z|a,b,z_p,\tau) = \int_{z_m}^{\infty}R_{\rm SFR}(z_f|a,b,z_p)P(z_m|z_f,\tau)\mathrm{d}z_f\,,
\label{delay_z}
\end{equation}
where \(\tau\) is the characteristic DTD. Let \(t_f\) and \(t_m\) denote the cosmic times corresponding to \(z_f\) and \(z_m\), respectively. Then \(P(z_m|z_f,\tau)\) can be written as \citep{vitale2019measuring}:
\begin{equation}
P(z_m|z_f,\tau) = {1\over\tau}\exp{\big[-\frac{t_f(z_f)-t_m(z_m)}{\tau}\big]}\frac{\mathrm{d}t}{\mathrm{d}z}\,.
\label{kernal}
\end{equation}

For the cosmic SFR, we apply the unnormalized Madau-Dickinson formalism \citep{madau2014cosmic}:
\begin{equation}
R_{\rm SFR}(z|a,b,z_p) = \frac{(1+z)^a}{1+\left(\frac{1+z}{1+z_p}\right)^{(a+b)}}\,.
\end{equation}
The physical interpretation is that, in the early Universe, gas was more abundant and galaxy mergers occurred more frequently, which made star formation easier, leading to a rise in the rate with index \(a\). At even earlier epochs, however, the star formation rate decreases with index \(b\) due to factors such as the lower masses of dark matter halos and the low metal abundance. The transition between these two regimes occurs near \(z=z_p\).

The corresponding population distribution function must be multiplied by the comoving volume element and account for cosmological time dilation, giving
\begin{equation}
p_{\rm pop}(z|\Vec{\Lambda}_z) = \mathcal{R}_n
\frac{R(z|\Vec{\Lambda}_z)}{1+z}\frac{\mathrm{d}V}{\mathrm{d}z}\,,
\end{equation}
where \(\mathcal{R}_n\) is a normalization factor related to the total population size.

\(\bullet\) \(k\)-correction factor \(k(z)\): The \(k\)-correction is closely related to the GRB spectrum and should, in principle, be described by a set of spectral hyperparameters \(\Vec{\Lambda}_k\). However, since this work does not focus on the spectral properties of GRBs, introducing additional hyperparameters would significantly increase the model complexity. Therefore, we directly assume that \(k(z)\) follows an empirical distribution \(p_{\rm pop}(k(z))\), chosen to match the distribution of \(k(z)\) derived from the current GBM catalog. In practice, after randomly generating a redshift \(z\) according to its population model, we randomly select a GRB event with a spectral fit from the existing sample and use its spectrum to compute the corresponding \(k\)-correction factor \(k(z)\) at the simulated redshift \(z\).

We further assume that the population distributions of the above parameters are mutually independent. The overall population distribution can therefore be written as:
\begin{equation}
\begin{aligned}
p_{\rm pop}(\Vec{\theta}|\Vec{\Lambda}) =
& p_{\rm pop}(\theta_c|\mu_{\theta_c},\sigma_{\theta_c})
p_{\rm pop}(L_0|A,L_*) \\
& p_{\rm pop}(z|a,b,z_p)
p_{\rm pop}(k(z))
p_{\rm pop}(\theta_v)\,.
\end{aligned}
\end{equation}
\phantom{x}

\begin{table}[]
\centering
\renewcommand{\arraystretch}{1.5}
\caption{Prior distributions for the hyperparameters \(\Vec{\Lambda}\): The prior choices are mainly based on S23, with modifications motivated by the GRB population model adopted in this work.}
\setlength{\tabcolsep}{42pt}
\begin{tabular}{cc}
\hline
\(\Lambda\)             & prior       \\ \hline
\(\mu_{\theta_c}\)    & \(U(0.03,0.2)\)    \\
\(\sigma_{\theta_c}\) & \(U(0.2,1.0)\) \\
\(A\)                 & \(U(1.5,5)\)    \\
\(\log{L_*}\)         & \(U(50.0,51.5)\)    \\
\(a\)                 & \(U(0,5)\)      \\
\(b\)                 & \(U(0,10)\)     \\
\(z_p\)               & \(U(0.1,5)\)    \\
\(\log\tau\)          & \(U(-2,2)\)     \\
\(\log{R_n}\)         & \(U(-1,5)\)    \\ \hline
\end{tabular}
\label{tab:prior}
\end{table}

\subsection{Algorithm}
\label{Algorithm}

In Equation~\ref{like_2}, both the numerator and the denominator involve multidimensional integrals with relatively complex integrands. To handle this, we approximate the integrals using the Monte Carlo method:
\begin{equation}
I = \int_a^b f(x)\mathrm{d}x \approx (b-a)\mathbb{E}[f(x)] = \frac{b-a}{N} \sum_{k=1}^{N} f(x_k)\,,
\end{equation}
where \(x\) is drawn from a uniform distribution \(x\sim U(a,b)\). If \(x\) is sampled from an arbitrary probability distribution \(p(x)\), the integral can be expressed as
\begin{equation}
I = \int_a^b \frac{f(x)}{p(x)}p(x)\mathrm{d}x \approx \frac{1}{N} \sum_{k=1}^{N} \frac{f(x_k)}{p(x_k)}\,.
\end{equation}
In the specific case of this work, since the redshift \(z\) and \(k\)-correction \(k(z)\) distributions may be delta functions, these two parameters are sampled directly from their corresponding likelihood function. For the remaining parameters, since the effective support of \(p(\Vec{d}|\Vec{\theta})\) may only partially overlap with that of \(p_{\rm pop}(\Vec{\theta}|\Vec{\Lambda})\), sampling from a uniform distribution is generally more efficient. In this work, for the integrals in Eq.~\ref{like_2}, we use \(N=10^5\) randomly generated samples. For the marginalization over \(\sigma_F\) in Eq.~\ref{eq:pdet}, we use \(N=10^4\) samples.

The integral in Eq.~\ref{delay_z} is not evaluated using the Monte Carlo method, since it contains only a single integration variable. The main difficulty is that, for each set of parameters \(\{a, b, z_p, \tau\}\), the corresponding event-rate integral must be recomputed, which would significantly increase the computational cost. To reduce the computational burden, we transform the integral into a convolution form for evaluation. The detailed derivation is presented in Appendix~\ref{fft}.

To infer the posterior distribution of \(\Vec{\Lambda}\), we employ the MCMC algorithm. In computing the likelihood function, GPU parallelization is utilized and calculations are accelerated via \texttt{jax} \citep{jax2018github}, resulting in an approximate 150-fold increase in efficiency. In order to more thoroughly explore the full prior space, we adopt the parallel-tempered MCMC sampler \texttt{eryn} \citep{Karnesis:2023ras,michael_katz_2023_7705496,2013PASP..125..306F}.

Specifically, we run 20 chains in parallel, each with 10 temperature levels, for a total of 4000 steps, retaining the last 2000 steps as stable posterior samples. The prior distributions of the parameters are set as listed in the Table \ref{tab:prior}. The analysis code developed for this work has been archived in an online repository \citep{cao2026software}.

\section{Results}

\begin{figure*}
    \centering
    \includegraphics[width=\linewidth]{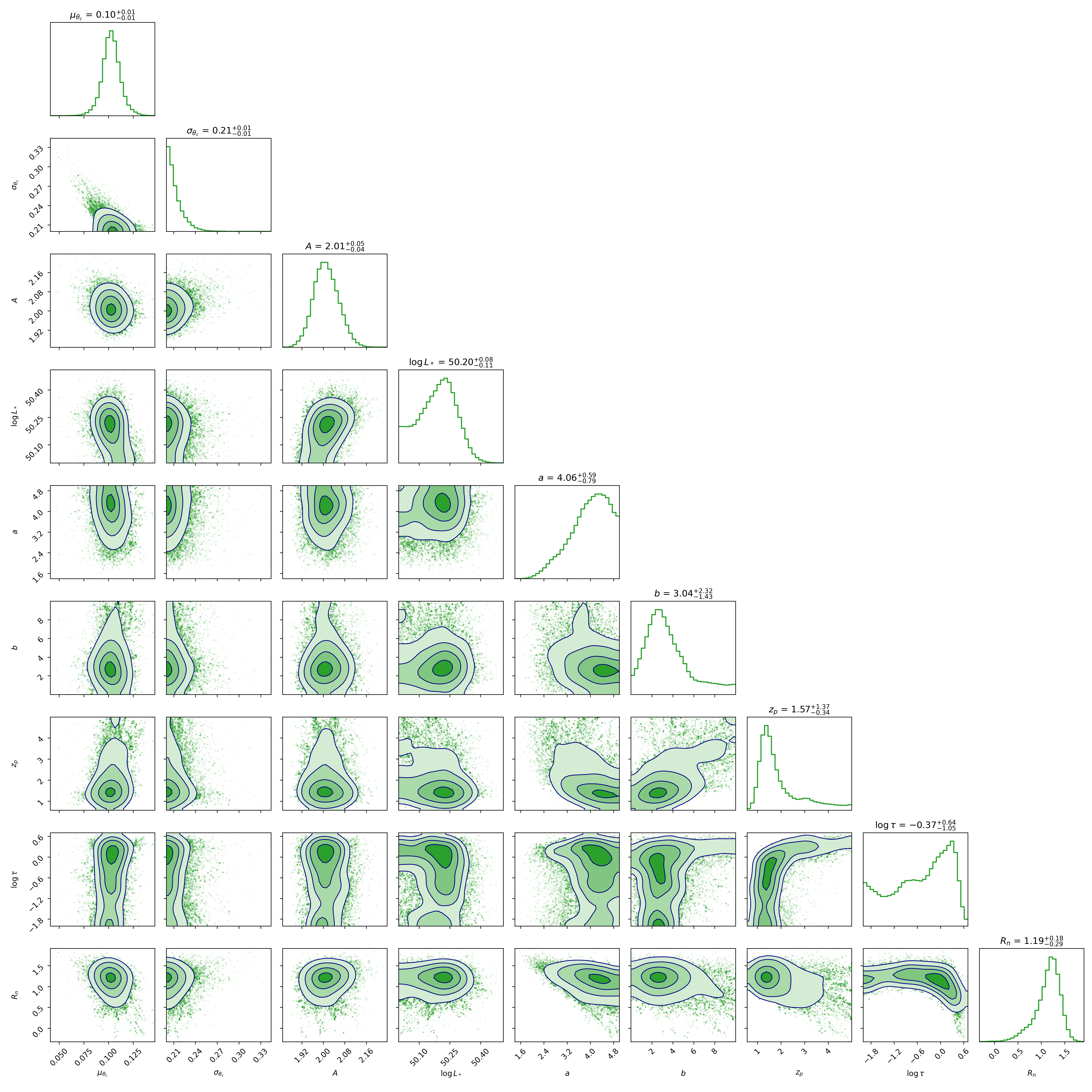}
    \caption{Corner plot of posterior distributions: The intervals labeled above each column correspond to \(1\sigma\) the posterior ranges.}
    \label{fig:linear_corner_noemf}
\end{figure*}

\begin{figure}
    \centering
    \includegraphics[width=\linewidth]{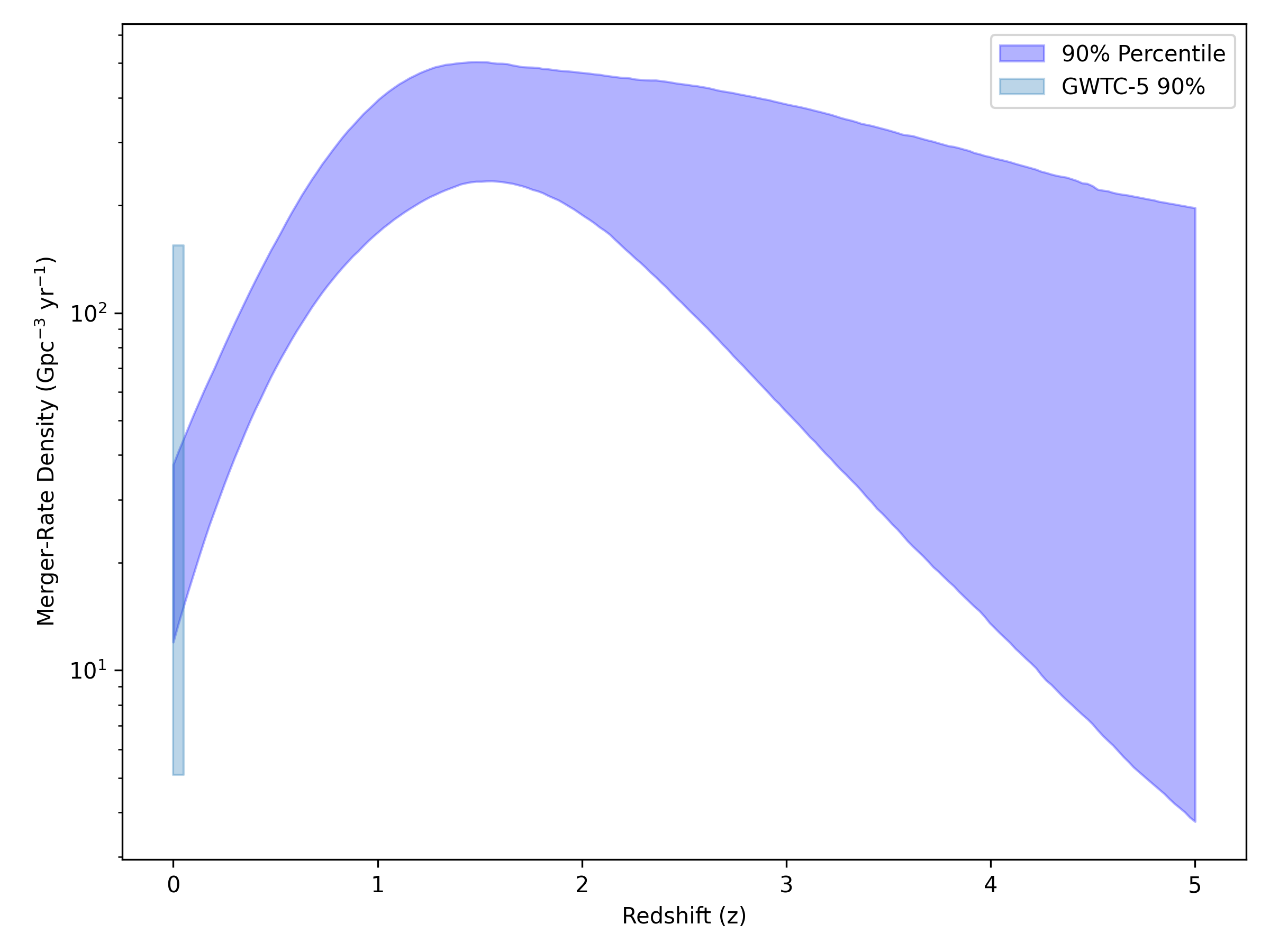}
    \caption{90\% credible intervals of the merger rate evolution with redshift. The distribution inferred from GWTC-5.0 \cite{ligo2026gwtc} is also shown for comparison.}
    \label{fig:linear_merge_rate_noemf}
\end{figure}

We present the posterior distributions in Figure~\ref{fig:linear_corner_noemf}. Overall, the jet opening angle and the on-axis luminosity are relatively well constrained. In contrast, because the available redshift information is limited, most of the parameters describing the redshift distribution remain only weakly constrained and show clear correlations.

For the jet opening angle, we infer a characteristic mean value of \(\theta_c \simeq 0.10\,{\rm rad}\). This value is larger than that reported by S23, but is close to the estimate of \cite{coward2012swift}. The model favors a relatively small dispersion in \(\theta_c\). However, owing to limitations of the present model and in order to maintain numerical stability, we do not further reduce this dispersion.

The lower cutoff of the on-axis luminosity distribution, \(L_*\), is constrained to be around \(1.6\times10^{50}\,{\rm erg\,s^{-1}}\), while the luminosity-distribution slope, \(-A\), is approximately 2. It should be noted that the on-axis luminosity \(L_0\) considered here is defined in the \(50\text{--}300\,\mathrm{keV}\) band and averaged over \(t_{90}\). Therefore, it is lower than the commonly quoted bolometric or peak on-axis luminosity. Similar to S23, we also obtain a lower constraint on $L_*$. Under a typical Band spectrum and FRED light curve, our definition of $L_*$ is approximately equivalent to the bolometric luminosity adopted in their work. Unlike their result, where the posterior distribution lies close to the lower edge of the prior range, our analysis yields a well-defined posterior peak at approximately $1.6\times10^{50}\,{\rm erg\,s^{-1}}$.

It is worth noting that the posterior still exhibits non-negligible structure below $10^{50}\,{\rm erg\,s^{-1}}$. However, lower luminosities imply a substantially larger fraction of undetected events, making the inferred distribution increasingly sensitive to selection effects and therefore less robust. Moreover, such low luminosities are not supported by conventional observational measurements \citep{coward2012swift, mooley2018superluminal}. For these reasons, we do not consider this low-luminosity regime in the following discussion.

The parameters controlling the redshift evolution are much less tightly constrained. The low-redshift power-law index of the SFR component is constrained to lie in the range \(a>3\), whereas the range of high-redshift power-law index is \(1.5<b<5.5\) . The peak redshift of the SFR, \(z_p\), is constrained to be between 1 and 3, and exhibits strong correlations with both \(a\) and \(b\). Specifically, as \(z_p\) increases, the inferred value of \(a\) becomes larger, while \(b\) becomes smaller. 

It is worth noting that the redshift reported by S23 corresponds to the peak of the GRB event-rate density. In our parametrization, however, \(z_p\) represents the peak redshift of the SFR. The DTD shifts the peak of the GRB event rate to lower redshift. Therefore, \(z_p\) and the characteristic delay time \(\tau\) jointly determine the final peak redshift of the GRB event-rate density, and consequently show a strong correlation. The characteristic DTD is constrained to lie within \(100\,{\rm Myr}<\tau<4\,{\rm Gyr}.\) As suggested by the \(z_p\)--\(\tau\) correlation shown in Figure~\ref{fig:linear_corner_noemf}, if one further assumes \(z_p>2\), the allowed range of the DTD can be narrowed to \(\tau>1\,{\rm Gyr}\). Conversely, a shorter DTD implies a lower value of $z_p$.

We further present the redshift evolution of the event-rate density in Figure~\ref{fig:linear_merge_rate_noemf}. To remain consistent with the observed number of detected events after accounting for selection effects, different treatments of the redshift evolution lead to different constraints on the normalization factor \(R_n\), and hence on the local event rate \(R_0\). The inferred evolutionary trend is broadly consistent with that of S23. However, since our calculation integrates over all energy ranges and all viewing angles, Figure~\ref{fig:linear_merge_rate_noemf} represents the intrinsic distribution of all Type I GRBs in the Universe. The local event rate is constrained to be \(R_0 = 22.6^{+15.0}_{-10.6}\,{\rm Gpc^{-3}\,yr^{-1}}.\)

The inferred local event rate is consistent with the latest constraints from gravitational-wave observations reported in GWTC-5.0 \citep{ligo2026gwtc}. Compared with previous estimates based on Type~I GRBs, our result lies toward the lower end of the range reported by S23, is comparable to the estimate of \cite{dominik2013double}, and is substantially lower than that of \cite{coward2012swift}. Assuming a typical jet opening angle, the estimates of \cite{wanderman2015rate} and \cite{ghirlanda2016short} are also consistent with our result. The difference from S23 mainly stems from two factors. First, rather than assuming a universal jet structure, we allow for intrinsic diversity in jet properties. Second, while our constraint on \(L_*\) is driven primarily by the likelihood, the posterior of S23 is prior-limited. The rate estimated by \citep{coward2012swift}, based on a non-Bayesian method, is dominated by a small number of nearby events, which biases the inferred viewing-angle distribution of higher-redshift sources toward smaller values. This, in turn, overestimates the beaming correction and consequently yields a higher local event rate.

\section{Discussions}
\label{sec:discussion}

For the jet opening angle, we assume that it follows a structured population distribution, rather than requiring all Type I GRBs to share a single value of the jet opening angle. However, as discussed in Appendix~\ref{app_4}, the dispersion of the jet opening angle cannot be set too small in order to maintain the numerical stability of the integration. As a result, the inferred dispersion of the jet opening angle lies close to the lower boundary of its prior. This result may indicate that the intrinsic dispersion is indeed very small, but that the present model does not allow us to constrain its exact value. Nevertheless, we cannot rule out the possibility that some of the model assumptions are inappropriate, such as the Gaussian jet model itself (i.e., Eq.~\ref{eq:L}) or the assumption that the jet opening angle follows a Gaussian distribution in log-space at the population level (i.e., Eq.~\ref{eq:theta_c_pop}).

For modeling Type I GRBs, obtaining more observational data to test the validity of theoretical models may be the most direct and effective approach. In terms of modeling parameter distributions, adopting non-parametric methods for parameters such as the jet opening angle may be more robust \citep{ray2023nonparametric,yi2022gravitational}. Compared to parametric models, such approaches avoid assuming a specific functional form and may therefore yield distributions closer to reality; however, this comes at the cost of significantly increased computational complexity and resource requirements.

Although BNS mergers are generally regarded as the leading progenitor channel for Type I GRBs, the inferred population may also include contributions from other channels. For example, NS--BH mergers can produce relativistic jets under favorable tidal-disruption conditions and would still belong to the broader compact-merger population \citep{narayan1992gamma,nakar2007short,berger2014short}. In addition, nearby magnetar giant flares may contaminate the low-luminosity, low-redshift end of the short-GRB sample \citep{hurley2005exceptionally}, while a small fraction of collapsar-origin bursts can enter a duration-selected short-GRB catalogue \citep{bromberg2013short}. These contaminants would affect the inferred rate evolution differently: collapsar contamination may bias the population toward a star-formation-like redshift evolution, whereas magnetar flares mainly influence the local low-luminosity rate. Therefore, our results should be interpreted as constraints on the Type I GRB population selected by the adopted criteria, rather than as a direct and exclusive measurement of the BNS merger rate.

The characteristic DTD of $\tau \approx 1$ Gyr inferred from our hierarchical Bayesian analysis of Type I GRBs is consistent with the double neutron star population observed in the Milky Way, after accounting for observational selection biases and subpopulation decomposition. \citet{2019ApJ...880L...8A} showed that the Galactic DNS sample decomposes into three distinct subpopulations; the short-period, high-eccentricity systems (subpopulation iii, including the Hulse-Taylor binary) have total evolutionary timescales from formation to merger peaking in the tens to hundreds of Myr range, overlapping with our measured \(100\,{\rm Myr}<\tau<4\,{\rm Gyr}.\) and corresponding to the fast-merging channel expected to produce short-duration GRBs with relativistic jets. This is supported by \citet{2004ApJ...601L.179K} and \citet{2019MNRAS.487.4847B}, who found that 40--60\% of all merging Galactic DNS systems fall into a rapid-merger category with delay times below 1 Gyr, while longer-period systems with Gyr-scale coalescence times contribute negligibly to the GRB sample. Independent support comes from Galactic chemical evolution: \citet{2019MNRAS.486.2896S} showed that reproducing the observed [Eu/Fe] abundance pattern and early r-process enrichment in metal-poor halo stars requires a substantial fraction of neutron star mergers to occur within the first few hundred Myr after star formation, consistent with our measured timescale. The apparent discrepancy with Gyr-scale merger times measured for some individual Galactic DNS systems (e.g., PSR J1913+1102) reflects a selection effect: radio pulsar surveys are biased toward longer-lived systems that spend more time in the observable radio-emitting phase, whereas our GRB population inference probes the intrinsic DTD of the jet-producing DNS subset directly.

The methodology presented here can also be applied to constrain the population properties of long gamma-ray bursts (lGRBs). For lGRBs, the time delay can be approximately neglected, and the Madau–Dickinson star formation rate can be directly used as a proxy for their merger rate. Although lGRBs differ from Type I GRBs in parameters such as isotropic-equivalent luminosity and jet opening angle, we assume that they share similar key parameters \(a, b, z_p\). Considering that {\it Fermi}/GBM has detected approximately five times more lGRBs than Type I GRBs \citep{von2020fourth,gruber2014fermi,von2014second,bhat2016third}, including many high-redshift sources \citep{barthelmy2005burst,lan2023grb,salafia2023short}, we expect tighter constraints on the parameter space.

When exploring the relation between peak photon flux \(F_{\rm ph}\) and energy flux \(F_{\rm E}\), we initially assumed a linear relation based on physical considerations. However, direct fitting results indicate that a linear relation does not fully capture the data, so we further consider a nonlinear relation \(F_{\rm ph}(F_{\rm E}, \sigma_F)\). The final results are presented in Appendix~\ref{app_3}. The result is broadly consistent with that obtained in the linear case, indicating that the final inference is largely insensitive to the assumed relation between $F_{\rm ph}$ and $(F_{\rm E}, \sigma_F)$.

We also attempted to supplement the missing redshift information using an empirical relation, as described in Appendix~\ref{emrs}. Although this approach significantly reduces the uncertainties of the redshift-distribution parameters, it appears to favor a solution that is difficult to reconcile with current theoretical expectations, which suggests that introducing redshift information through EMR may lead to a systematic underestimation of the inferred redshifts. A reasonable reason is, in this work, we use GRBs with known redshifts to infer the distribution of events with unknown redshifts. However, GRBs with measured redshifts tend to be biased toward low-redshift or high-flux events, inevitably introducing selection effects. Such biases can propagate into the EMR-calibrated pseudo-redshift distribution and hence affect the inferred luminosity function, selection correction, and rate normalization.

Moreover, some studies have pointed out that pseudo-redshifts estimated from EMRs may carry large uncertainties \citep{yorgancioglu2025feasibility,yorgancioglu2025can}, and it has even been argued that these EMRs may not reflect intrinsic physical correlations, but instead arise as apparent correlations induced by selection effects \citep{collazzi2012significant,huang2021reconciling}. Therefore, the discrepancy between the EMR-based and non-EMR-based rate estimates should be interpreted as a systematic uncertainty associated with the treatment of redshift information, rather than as evidence for two distinct intrinsic rate histories.

As discussed in Section~\ref{sect:HB}, a more rigorous treatment of selection effects requires generating a large number of simulated sources, adding realistic noise, and evaluating their detectability \citep{abbott2023population,abac2025gwtc,ligo2026gwtc}. Our next step is to construct such a simulated source catalog to systematically constrain the hyperparameters of the population. In addition to the {\it Fermi}/GBM catalog, other detectors such as {\it Swift}/BAT provide substantial GRB observations. By generating simulated catalogs that account for multiple detectors, we expect to further tighten the constraints on population parameters and enhance the robustness of our results.

It is worth noting that, as discussed in Section~\ref{Algorithm}, we approximate the integrals using Monte Carlo methods, and thus the associated statistical uncertainty must be taken into account. The derivation of this uncertainty is provided in the Appendix \ref{app_4}. In principle, this uncertainty can be reduced to an arbitrarily small level given a sufficiently large number of samples. However, for the Type I GRB population model, the distributions—particularly for the jet opening angle \((\theta_c)\) and on-axis luminosity \((L_0)\)—are highly concentrated. Compared to the relatively broader population distributions in gravitational-wave events, significantly more samples are required to achieve the same level of precision. Under current computational resources, it is nearly infeasible to reach the same convergence criteria as in \cite{ligo2026gwtc}.

In \cite{mancarella2025sampling}, the authors treat the source parameters of each event as free parameters to be sampled jointly, thereby avoiding the stochastic error introduced by the integral in the numerator of Eq.~\ref{like_2}, such that the overall uncertainty is dominated by the selection effects. However, this approach places higher demands on the performance and efficiency of the sampler. The uncertainty introduced by selection effects can be significantly reduced through large-scale simulated population injections, which is also a direction we plan to pursue in future work.  While this paper was in preparation, \cite{2026ApJ..1006L..49F} discussed a possible tension between the short-GRB rate and the GW-inferred BNS merger rate; our fully hierarchical Bayesian analysis, by contrast, yields a local Type I GRB rate consistent with the GW-inferred value.

\begin{acknowledgments}
This work was supported by the National Natural Science Foundation of China (NOs. 12494572, 12494570), the Strategic Priority Research Program of the Chinese Academy of Sciences (Grant No. XDB0550300), and China’s Space Origins Exploration Program.
The authors are grateful for the public data of Fermi/GBM.
\end{acknowledgments}

\begin{contribution}

All authors contributed equally collaboration.


\end{contribution}

%



\appendix
\section{the convolution form of the time-delay model}
\label{fft}

We first recall the definition of convolution:
\begin{equation}
(f*g)(x) = \int_{-\infty}^{\infty} f(y)g(x-y)\mathrm{d}y\,.
\end{equation}
In numerical calculations, this integral can be discretized as a discrete convolution:
\begin{equation}
(f*g)[i] = \sum_{j=0}^{N-1} f[j] g[i-j]\,.
\end{equation}
This form can be efficiently computed on GPUs using fast convolution algorithms such as the FFT.

Substituting Eq.~\ref{kernal} into Eq.~\ref{delay_z} and transforming the integration variable from redshift \(z\) to the lookback time \(t_f\), we obtain:
\begin{equation}
R(t_m|\Vec{\Lambda}) =
\int_{t_m}^{\infty}
R_{\rm SFR}(t_f|a,b,z_p)
\frac{1}{\tau}
\exp\left[-\frac{t_f-t_m}{\tau}\right]
\mathrm{d}t_f\,.
\end{equation}

The form of this integral is similar to a convolution, but its integration range is \([t_m,\infty)\) rather than the standard convolution range \((-\infty,\infty)\). To transform it into a convolution form, we perform the variable substitution \(\Delta t = t_f - t_m\) so that the integral becomes:
\begin{equation}
\begin{aligned}
R(t_m|\Vec{\Lambda})
&=
\int_{0}^{\infty}
R_{\rm SFR}(t_m+\Delta t|a,b,z_p)
\frac{1}{\tau}
\exp\left[-\frac{\Delta t}{\tau}\right]
\mathrm{d}\Delta t \\
&=
\int_{-\infty}^{0}
\frac{1}{\tau}
\exp\left[\frac{s}{\tau}\right]
R_{\rm SFR}(t_m-s|a,b,z_p)
\mathrm{d}s\,,
\end{aligned}
\end{equation}
where \(s=-\Delta t\). It can be seen that the integration range is now \((-\infty,0]\), and thus the expression can be regarded as a semi-infinite convolution.

To rewrite it in the standard convolution form, we define:
\begin{equation}
\begin{aligned}
f(t) &=
\begin{cases}
\frac{1}{\tau}\exp\left(\frac{t}{\tau}\right), & t<0 \\
0, & t>0
\end{cases}\,, \\
g(t) &= R_{\rm SFR}(t|a,b,z_p)\,.
\end{aligned}
\end{equation}
Since \(f(t)=0\) for \(t>0\), the integration range can be naturally extended to the entire real axis, yielding the standard convolution form:
\begin{equation}
R(t|\Vec{\Lambda}) =
\int_{-\infty}^{\infty} f(s) g(t-s)\mathrm{d}s
= (f*g)(t)\,.
\end{equation}

In practical calculations, after discretizing time, the sequence \(g[i]\) corresponds to the discrete values of the function \(R_{\rm SFR}(t|a,b,z_p)\) arranged in chronological order, while the sequence \(f[i]\) corresponds to the reversed discrete sequence of the function \(\frac{1}{\tau}\exp\left(-\frac{t}{\tau}\right)\). Therefore, the problem can be efficiently solved using discrete convolution.

\section{the result using nonlinear relation \(F_{\rm ph}(F_{\rm E}, \sigma_F)\)}
\label{app_3}

\begin{figure*}
    \centering
    \includegraphics[width=.8\linewidth]{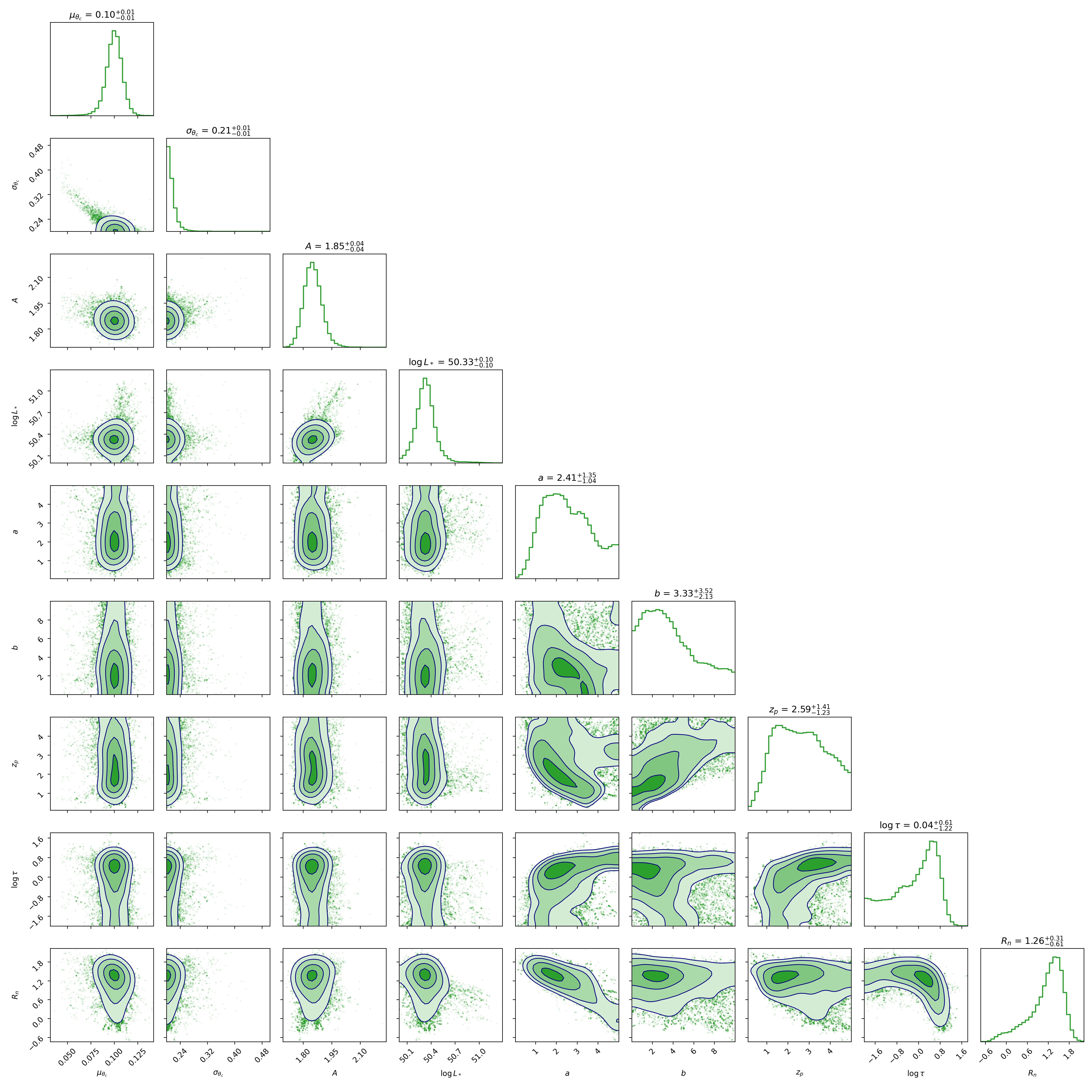}
    \caption{Corner plot of posterior distributions but with nonlinear relation \(F_{\rm ph}(F_{\rm E}, \sigma_F)\)}
    \label{fig:poly_corner}
\end{figure*}

The corner plot of the posterior distribution obtained using the nonlinear relation \(F_{\rm ph}(F_{\rm E}, \sigma_F)\), along with the redshift evolution of the merger rate and the distribution of the local merger rate, are presented in this section, shown as Fig~\ref{fig:poly_corner}. The adopted nonlinear relation and its best-fitting parameters are shown in Figure~\ref{fig:flux_fit} and summarized in Table~\ref{tab:flux_fit}, respectively.

The posterior distributions of the hyperparameters obtained using the nonlinear relation are broadly consistent with those derived from the linear relation, indicating that the final inference is largely insensitive to the assumed relation. Specifically, the characteristic jet opening angle decreases to $\theta_c\simeq0.1\,{\rm rad}$, while the characteristic luminosity increases to $L_*\simeq2\times10^{51}\,{\rm erg\,s^{-1}}$. This trend further illustrates the correlation between these two parameters, as a narrower jet requires a higher intrinsic luminosity to produce a comparable observed flux. Meanwhile, the luminosity-distribution slope, $-A$, decreases to 1.85, implying that the luminosity distribution becomes more concentrated around $L_*$. For the redshift distribution, the main differences are a smaller value of $a$ and a larger value of $z_p$, indicating that a larger fraction of sources are distributed at higher redshifts and that the merger-rate evolution rises more gradually with redshift. The corresponding local merger rate is inferred to be \(R_0 = 32.3^{+27.8}_{-21.0}\,{\rm Gpc^{-3}\,yr^{-1}}\).

\begin{figure}[H]
    \centering
    \includegraphics[width=.8\linewidth]{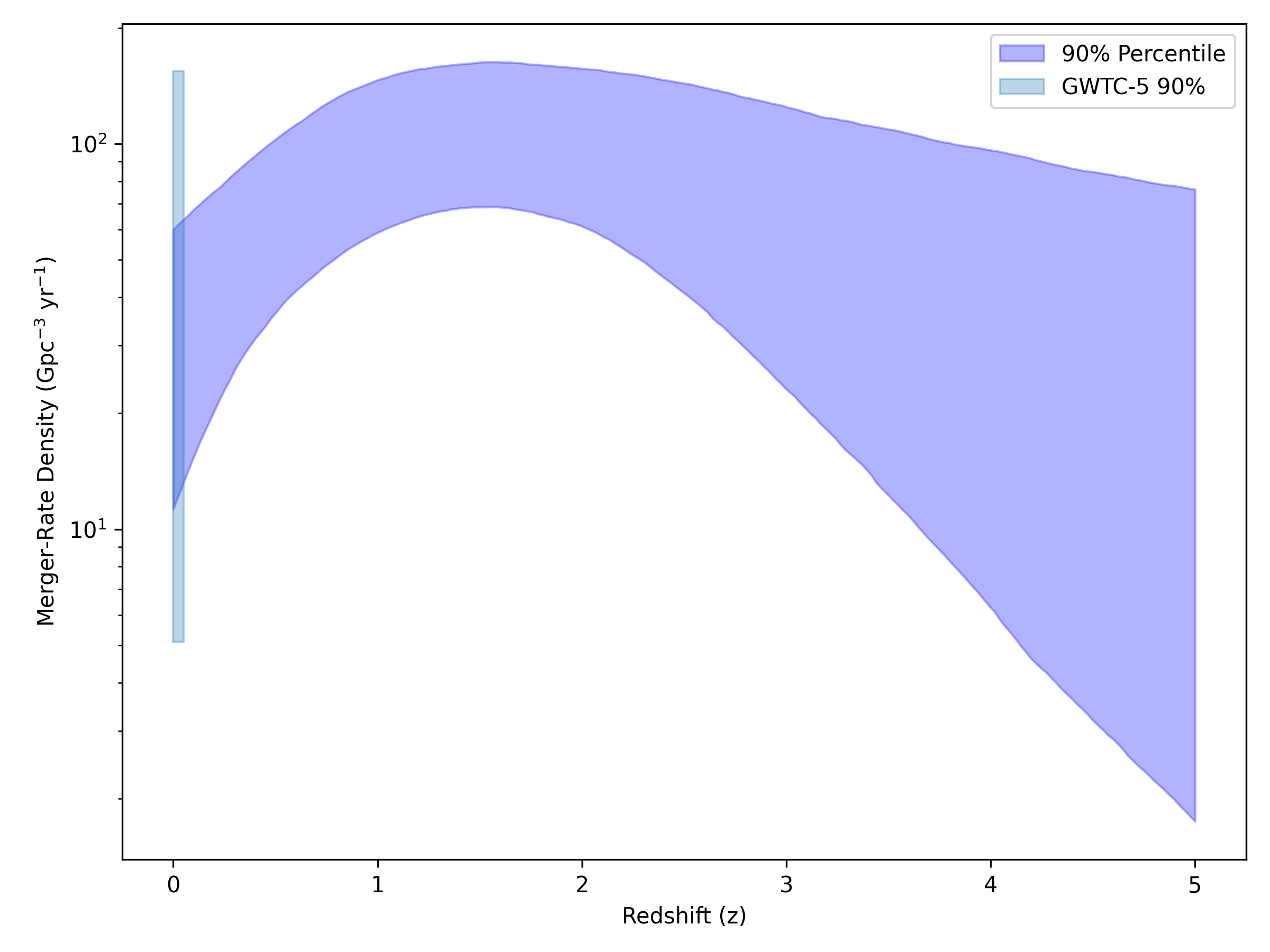}
    \caption{90\% credible intervals of the merger rate evolution with redshift, but with nonlinear relation \(F_{\rm ph}(F_{\rm E}, \sigma_F)\)}
    \label{fig:poly_merge_rate}
\end{figure}

\clearpage
\section{empirical relationships}
\label{emrs}

Utilising existing redshift measurements and empirical relationships, it is possible to supplement the missing redshift information. In particular, rather than directly adopting the analytical forms of the Amati and Yonetoku relations \citep{amati2002intrinsic, yonetoku2004gamma}, we construct the covariance matrix among \(\log E_{\rm iso}\), \(\log\!\big[t_{90}/(1+z)\big]\), and \(\log\!\big[E_{\rm p}(1+z)\big]\) using the subsample with known redshifts. In this way, the intrinsic scatter and mutual correlations of these observables are consistently taken into account.

Using the subsample with measured redshift, we estimate the covariance matrix \(\Tilde{C}\) for Type I GRBs. For events with spectral data, {\it Fermi}/GBM provides \(\Tilde{E}_{\rm p}\) from Band-function fitting, while \(\Tilde{E}_{\rm iso}\) is computed as:
\begin{equation}
\Tilde{E}_{\rm iso} =
\frac{4\pi D_L^2(\Tilde{z})\, k(\Tilde{z})\, \Tilde{f}_{\rm E}}{1+\Tilde{z}}\,,
\end{equation}
where \(D_L(\Tilde{z})\) is the luminosity distance at redshift \(\Tilde{z}\).

Combining these quantities with the observed \(\Tilde{t}_{90}\), the EMRs can be used to construct a likelihood for the redshift:
\begin{equation}
\mathcal{L}(z \mid \Tilde{E}_{\rm iso}, \Tilde{E}_{\rm p}, \Tilde{t}_{90}) \propto
\exp\left(-\frac{1}{2} \Delta \Vec{x}\cdot \Tilde{C}^{-1}\cdot \Delta \Vec{x}^{\mathrm{T}}\right)\,,
\end{equation}
where
\begin{equation}
\Vec{x} = \left(\log \Tilde{E}_{\rm iso}\,, \log \frac{\Tilde{t}_{90}}{1+z}\,, \log \big[\Tilde{E}_{\rm p}(1+z)\big]\right)\,,
\end{equation}
\(\Delta \Vec{x} = \Vec{x} - \Vec{x}_0\), and \(\Vec{x}_0\) denotes the mean vector derived from the redshift-known sample.

\begin{figure*}
    \centering
    \includegraphics[width=.8\linewidth]{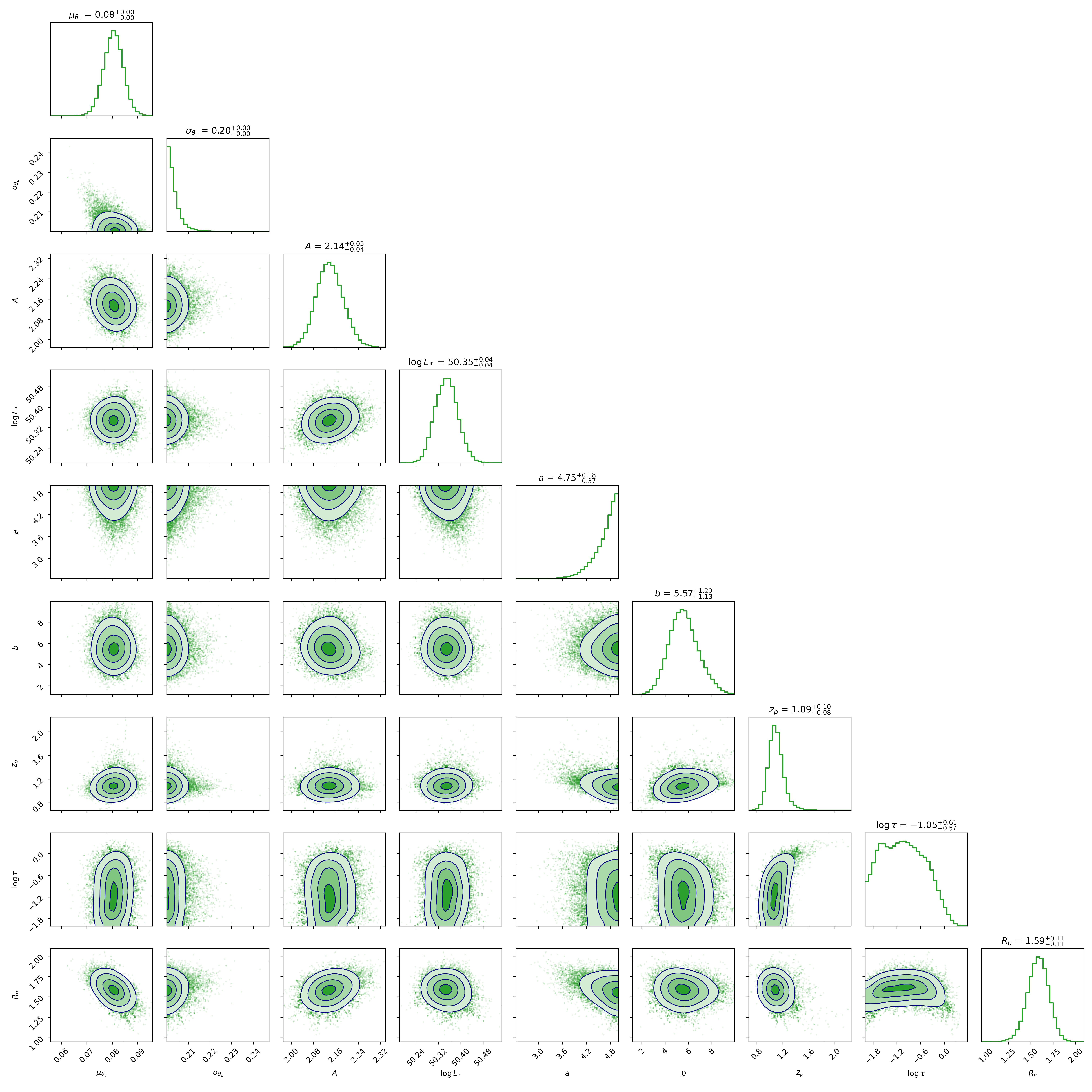}
    \caption{Corner plot of posterior distributions with EMRs}
    \label{fig:linear_corner_noemr}
\end{figure*}

By incorporating redshift information using the method described above, and assuming a linear \(F_{\rm ph}(F_{\rm E}, \sigma_F)\) relation, we obtain the posterior distributions of the hyperparameters shown in Figure~\ref{fig:linear_corner_noemr}. As expected, the inclusion of the additional redshift information significantly improves the constraints on the parameters governing the redshift distribution, namely \(a\), \(b\), \(z_p\), and \(\tau\).

\begin{figure}[H]
    \centering
    \includegraphics[width=.8\linewidth]{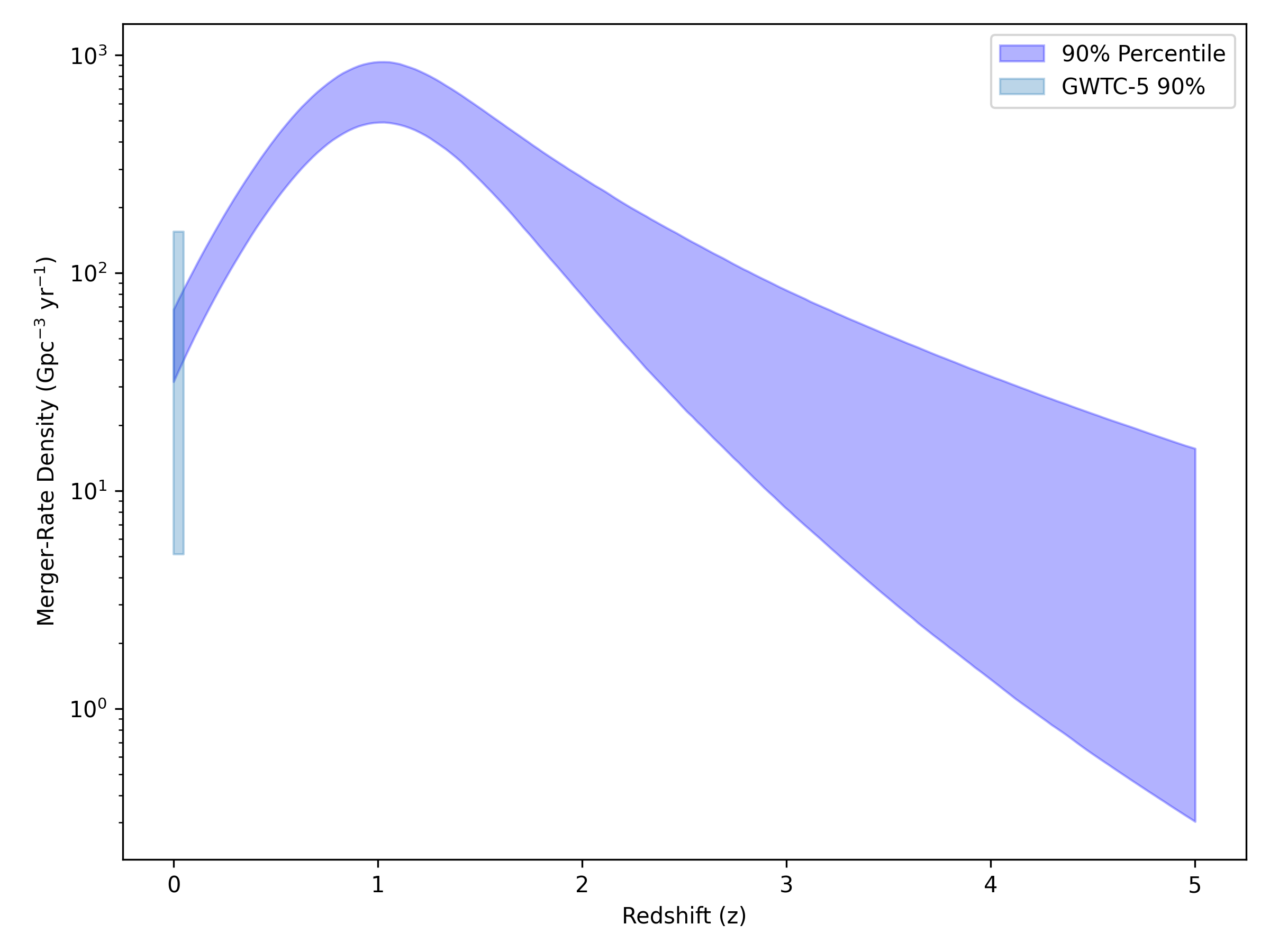}
    \caption{90\% credible intervals of the merger rate evolution with redshift with EMRs}
    \label{fig:linear_merge_rate_noemr}
\end{figure}

However, despite these tighter constraints, the inferred parameter values differ markedly from those obtained without adopting the empirical relation. In particular, the parameter \(a\) is driven toward the upper boundary of its prior, while \(z_p\) is constrained to a very low value. The corresponding merger-rate density is shown in Figure~\ref{fig:linear_merge_rate_noemr}. Overall, the inferred merger-rate distribution is strongly biased toward low redshifts. Nevertheless, owing to the assumed SFR model and the prior imposed on \(a\), the model is unable to produce an extremely concentrated low-redshift distribution.

\clearpage
\section{Monte Carlo uncertainty}
\label{app_4}

When estimating integrals using the Monte Carlo method, one typically samples the variable \(x\) from a normalized distribution \(p(x)\) (satisfying \(\int \mathrm{d}x p(x) = 1\)) and approximates the integral by the arithmetic mean of the samples, i.e.,
\begin{equation}
\int \mathrm{d}x\,f(x)p(x) = \langle f \rangle \approx \frac{1}{N}\sum_{k=1}^{N} f(x_k)\,.
\end{equation}
The variance of this estimator is:
\begin{equation}
\sigma^2_{\langle f \rangle} = \frac{1}{N}\left(\langle f^2 \rangle - \langle f \rangle^2\right)\,.
\end{equation}

In this work, we denote the integral in the numerator of Eq.~\ref{like_2} as \(\mathcal{N}\) and that in the denominator as \(\mathcal{D}\). The relative variance introduced by the likelihood of a single event can then be written as:
\begin{equation}
\sigma_i^2 =
\frac{\sigma^2_{\mathcal{N}_i}}{\langle \mathcal{N}_i \rangle^2}
+
\frac{\sigma^2_{\mathcal{D}}}{\langle \mathcal{D} \rangle^2}\,.
\end{equation}
For \(N_{\rm obs}\) observed events, the variance of the total log-likelihood is
\begin{equation}
\sigma^2(\ln \mathcal{L}) =
\sum_{i=1}^{N_{\rm obs}}
\frac{\sigma^2_{\mathcal{N}_i}}{\langle \mathcal{N}_i \rangle^2}
+
N_{\rm obs}^2
\frac{\sigma^2_{\mathcal{D}}}{\langle \mathcal{D} \rangle^2}\,.
\end{equation}

In GWTC-3 \cite{abbott2023population}, it is typically required that the effective sample size satisfies \(N_{\rm eff} > 4N_{\rm obs}\) to ensure the stability of the integral, where \cite{farr2019accuracy}
\begin{equation}
N_{\rm eff} = N_{\rm obs}\frac{\langle f \rangle^2}{\langle f^2 \rangle}\,.
\end{equation}
In GWTC-4.0 and GWTC-5.0 \cite{abac2025gwtc, ligo2026gwtc}, the stability criterion is further reformulated as \(\sigma^2(\ln \mathcal{L}) < 1\). Essentially, both criteria aim to constrain the statistical variance introduced by Monte Carlo integration. For some models, this requirement can significantly restrict the range of the hyperparameter space that can be explored, and therefore improving this algorithm remains an important problem in current research \citep{wysocki2019reconstructing,doctor2020black,delfavero2021normal,golomb2022hierarchical,mould2024calibrating,hussain2026hints,mancarella2025sampling}.

\section{Summary table of the properties of the sources used}
\label{app_5}

The measurements of the 599 Type I GRB events used in this work are categorized according to their best-fit spectral models and presented in five tables in this section.

\begin{longtable}{ccccccccc}
\caption{Sources with COMP classification: The units of quantities listed are \(t_{90}\) (s), \(f_{\rm E}\) (erg\,cm\(^{-2}\)), \(\Delta f_{\rm E}\) (erg\,cm\(^{-2}\)), \(F_{\rm ph}\) (photon\,cm\(^{-2}\)\,s\(^{-1}\)), \(\Delta F_{\rm ph}\) (photon\,cm\(^{-2}\)\,s\(^{-1}\)), \(A\) (photon\,cm\(^{-2}\)\,s\(^{-1}\)\,keV\(^{-1}\)), and \(E_{p}\) (keV), while \(\alpha\) is dimensionless.}\\
\toprule
name         & \(t_{90}\)   & \(f_{\rm E}\) & \(\Delta f_{\rm E}\) & \(F_{\rm ph}\) & \(\Delta F_{\rm ph}\) & \(A\) & \(E_{p}\) & \(\alpha\) \\ \midrule \endfirsthead

\toprule
name         & \(t_{90}\)   & \(f_{\rm E}\) & \(\Delta f_{\rm E}\) & \(F_{\rm ph}\) & \(\Delta F_{\rm ph}\) & \(A\) & \(E_{p}\) & \(\alpha\) \\ \midrule \endhead

\midrule
\endfoot

\bottomrule
\endlastfoot

GRB181222841 & 0.576 & 0.000017833    & 1.1011E-08            & 370.312         & 5.8218                 & 0.2472248        & 390.49            & -0.8185901        \\
GRB171108656 & 0.032 & 6.3508E-06     & 1.2387E-07            & 208.061         & 20.8367                & 0.05954138       & 100.2242          & -0.9947428        \\
GRB150819440 & 0.96  & 3.2965E-06     & 9.2123E-09            & 79.3628         & 2.7382                 & 0.1148794        & 523.8768          & -1.136249         \\
GRB180703949 & 1.536 & 5.6463E-06     & 9.7284E-09            & 61.117          & 1.9335                 & 0.2183298        & 136.8747          & -0.7720429        \\
GRB090228204 & 0.448 & 0.000001755    & 1.3201E-08            & 60.9947         & 2.186                  & 0.08311341       & 861.4447          & -0.6096338        \\
GRB160822672 & 0.04  & 1.2631E-06     & 7.4241E-08            & 50.4184         & 7.3119                 & 0.06866003       & 661.2325          & -0.822139         \\
GRB190606080 & 0.224 & 1.4418E-06     & 6.6572E-09            & 47.693          & 1.8368                 & 0.09250587       & 2344.394          & -1.055922         \\
GRB201227635 & 0.112 & 0.000000878    & 1.5551E-08            & 46.9079         & 3.2588                 & 0.1224955        & 1022.463          & -0.4876436        \\
\end{longtable}

\begin{longtable}{ccccccccc}
\caption{Sources with PLAW classification: The units of quantities listed are \(t_{90}\) (s), \(f_{\rm E}\) (erg\,cm\(^{-2}\)), \(\Delta f_{\rm E}\) (erg\,cm\(^{-2}\)), \(F_{\rm ph}\) (photon\,cm\(^{-2}\)\,s\(^{-1}\)), \(\Delta F_{\rm ph}\) (photon\,cm\(^{-2}\)\,s\(^{-1}\)), \(A\) (photon\,cm\(^{-2}\)\,s\(^{-1}\)\,keV\(^{-1}\)), while \(\alpha\) is dimensionless.}\\
\toprule
name         & \(t_{90}\)   & \(f_{\rm E}\) & \(\Delta f_{\rm E}\) & \(F_{\rm ph}\) & \(\Delta F_{\rm ph}\) & \(A\) & \(\alpha\) \\ \midrule \endfirsthead

\toprule
name         & \(t_{90}\)   & \(f_{\rm E}\) & \(\Delta f_{\rm E}\) & \(F_{\rm ph}\) & \(\Delta F_{\rm ph}\) & \(A\) & \(\alpha\) \\ \midrule \endhead

\midrule
\endfoot

\bottomrule
\endlastfoot

GRB211104782 & 0.704 & 9.8522E-07     & 2.2674E-07            & 31.2947         & 11.7039                & 0.01054689       & -1.141499         \\
GRB171223818 & 0.384 & 5.6913E-07     & 1.1692E-08            & 11.4077         & 0.8596                 & 0.0282346        & -1.247638         \\
GRB100805300 & 0.064 & 1.1064E-07     & 4.103E-08             & 11.3504         & 1.7842                 & 0.04626818       & -1.501964         \\
GRB180311074 & 0.96  & 4.1688E-07     & 1.8531E-08            & 7.6963          & 1.263                  & 0.01893023       & -1.441067         \\
GRB150128624 & 0.096 & 1.1324E-07     & 2.7749E-08            & 7.5819          & 1.2817                 & 0.02353447       & -1.596424         \\
GRB130808253 & 0.256 & 1.2241E-07     & 9.3852E-09            & 6.7661          & 0.8581                 & 0.01695146       & -1.855503         \\
GRB100107074 & 0.576 & 9.0539E-08     & 1.0998E-08            & 6.4307          & 1.4885                 & 0.01467815       & -1.243538         \\
GRB110801335 & 0.384 & 2.167E-07      & 2.0165E-08            & 6.1003          & 1.2103                 & 0.01750808       & -1.659831         \\
\end{longtable}

\begin{longtable}{cccccccccc}
\caption{Sources with SBPL classification: The units of quantities listed are \(t_{90}\) (s), \(f_{\rm E}\) (erg\,cm\(^{-2}\)), \(\Delta f_{\rm E}\) (erg\,cm\(^{-2}\)), \(F_{\rm ph}\) (photon\,cm\(^{-2}\)\,s\(^{-1}\)), \(\Delta F_{\rm ph}\) (photon\,cm\(^{-2}\)\,s\(^{-1}\)), \(A\) (photon\,cm\(^{-2}\)\,s\(^{-1}\)\,keV\(^{-1}\)), and \(E_{b}\) (keV), while \(\alpha, \beta\) is dimensionless.}\\
\toprule
name         & \(t_{90}\)   & \(f_{\rm E}\) & \(\Delta f_{\rm E}\) & \(F_{\rm ph}\) & \(\Delta F_{\rm ph}\) & \(A\) & \(\alpha\) & \(E_{\rm b}\) & \(\beta\) \\ \midrule \endfirsthead

\toprule
name         & \(t_{90}\)   & \(f_{\rm E}\) & \(\Delta f_{\rm E}\) & \(F_{\rm ph}\) & \(\Delta F_{\rm ph}\) & \(A\) & \(\alpha\) & \(E_{\rm b}\) & \(\beta\) \\ \midrule \endhead

\midrule
\endfoot

\bottomrule
\endlastfoot

GRB120323507 & 0.384 & 5.3681E-06     & 1.6868E-08            & 188.577         & 3.1859                 & 0.1850748        & -1.173803         & 42.16719          & -2.081979         \\
GRB171126235 & 1.472 & 4.8423E-06     & 2.6839E-08            & 60.9514         & 3.5562                 & 0.03103142       & -0.6794954        & 59.88982          & -2.649991         \\
GRB120624309 & 0.64  & 2.5545E-06     & 4.1883E-08            & 48.9318         & 1.3144                 & 0.05093549       & -0.938021         & 2988.733          & -2.985605         \\
GRB090227772 & 0.304 & 1.8133E-06     & 6.2395E-09            & 38.4254         & 1.4812                 & 0.05109121       & -0.6472595        & 1697.794          & -3.329988         \\
GRB170708046 & 0.144 & 4.6872E-07     & 6.1874E-09            & 25.7074         & 1.0165                 & 0.05455541       & -0.4641976        & 48.23826          & -1.816086         \\
GRB101208498 & 2.048 & 2.2189E-06     & 2.8157E-08            & 22.1601         & 1.159                  & 0.01260504       & -1.173174         & 39.2868           & -2.131442         \\
\end{longtable}

\begin{longtable}{cccccccccc}
\caption{
Sources with BAND classification: The units of quantities listed are \(t_{90}\) (s), \(f_{\rm E}\) (erg\,cm\(^{-2}\)), \(\Delta f_{\rm E}\) (erg\,cm\(^{-2}\)), \(F_{\rm ph}\) (photon\,cm\(^{-2}\)\,s\(^{-1}\)), \(\Delta F_{\rm ph}\) (photon\,cm\(^{-2}\)\,s\(^{-1}\)), \(A\) (photon\,cm\(^{-2}\)\,s\(^{-1}\)\,keV\(^{-1}\)), and \(E_{p}\) (keV), while \(\alpha, \beta\) is dimensionless.}\\
\toprule
name         & \(t_{90}\)   & \(f_{\rm E}\) & \(\Delta f_{\rm E}\) & \(F_{\rm ph}\) & \(\Delta F_{\rm ph}\) & \(A\) & \(E_{\rm p}\) & \(\alpha\) & \(\beta\) \\ \midrule \endfirsthead

\toprule
name         & \(t_{90}\)   & \(f_{\rm E}\) & \(\Delta f_{\rm E}\) & \(F_{\rm ph}\) & \(\Delta F_{\rm ph}\) & \(A\) & \(E_{\rm p}\) & \(\alpha\) & \(\beta\) \\ \midrule \endhead

\midrule
\endfoot

\bottomrule
\endlastfoot

GRB140209313 & 1.408 & 5.7305E-06     & 2.8082E-08            & 66.725          & 2.1842                 & 0.1914856        & 143.9007          & -0.6056056        & -2.399425        \\
GRB170206453 & 1.168 & 5.1375E-06     & 6.2294E-08            & 33.0444         & 1.4497                 & 0.1096833        & 347.9916          & -0.4577412        & -2.586704        \\
GRB200826187 & 1.136 & 2.7077E-06     & 1.2281E-08            & 30.0007         & 1.0441                 & 0.7566026        & 68.93214          & -0.05721819       & -2.0001          \\
GRB110529034 & 0.512 & 7.9636E-07     & 8.346E-09             & 22.459          & 0.9765                 & 0.03797514       & 990.4448          & -0.8673255        & -2.004221       
\end{longtable}

\begin{longtable}{cccccc}
\caption{
Sources without spectral fits: The units of quantities listed are \(t_{90}\) (s), \(f_{\rm E}\) (erg\,cm\(^{-2}\)), \(\Delta f_{\rm E}\) (erg\,cm\(^{-2}\)), \(F_{\rm ph}\) (photon\,cm\(^{-2}\)\,s\(^{-1}\)), \(\Delta F_{\rm ph}\) (photon\,cm\(^{-2}\)\,s\(^{-1}\)).}\\
\toprule
name         & \(t_{90}\)   & \(f_{\rm E}\) & \(\Delta f_{\rm E}\) & \(F_{\rm ph}\) & \(\Delta F_{\rm ph}\) \\ \midrule \endfirsthead

\toprule
name         & \(t_{90}\)   & \(f_{\rm E}\) & \(\Delta f_{\rm E}\) & \(F_{\rm ph}\) & \(\Delta F_{\rm ph}\) \\ \midrule \endhead

\midrule
\endfoot

\bottomrule
\endlastfoot

GRB230217912 & 0.896 & 4.7542E-06     & 2.1422E-08            & 39.9594         & 1.9528                 \\
GRB230306673 & 0.64  & 0.000000306    & 8.4769E-08            & 10.7148         & 6.5289                 \\
GRB190118937 & 0.576 & 3.1701E-07     & 4.6658E-08            & 7.4924          & 1.6614                 \\
GRB230419568 & 0.256 & 1.7718E-07     & 8.3627E-09            & 6.8343          & 1.0142                 \\
GRB140616165 & 0.512 & 1.847E-07      & 2.2422E-08            & 6.2925          & 1.0056                 \\
GRB150226223 & 1.408 & 6.1772E-07     & 1.4912E-08            & 6.1799          & 0.4996                 \\
GRB081229675 & 0.032 & 6.9932E-08     & 1.5229E-08            & 5.3717          & 0.7475                 \\
GRB220323507 & 0.448 & 1.1684E-07     & 1.141E-08             & 4.7989          & 0.734                  \\
\end{longtable}


\bibliography{sample701}{}

@software{cao2026software,
  author       = {Cao, T.-Y. and Yi, S.-X. and Du, Y.-F. and Yorgancioglu, E. S. and Xiong, S.},
  title        = {Software for Hierarchical Bayesian Inference on the Intrinsic Event Rate of Type I Gamma-Ray Bursts from the Fermi/GBM Catalog},
  year         = {2026},
  publisher    = {Zenodo},
  doi          = {10.5281/zenodo.22074410},
  url          = {https://doi.org/10.5281/zenodo.22074410}
}

@article{howell2025apparent,
  title={The apparent and cosmic event rate densities of short gamma-ray bursts},
  author={Howell, EJ and Burns, E and Goldstein, A},
  journal={Monthly Notices of the Royal Astronomical Society},
  volume={544},
  number={4},
  pages={3158--3172},
  year={2025},
  publisher={Oxford University Press}
}

@article{essick2024ensuring,
  title={Ensuring consistency between noise and detection in hierarchical Bayesian inference},
  author={Essick, Reed and Fishbach, Maya},
  journal={The Astrophysical Journal},
  volume={962},
  number={2},
  pages={169},
  year={2024},
  publisher={The American Astronomical Society}
}

@incollection{vitale2022inferring,
  title={Inferring the properties of a population of compact binaries in presence of selection effects},
  author={Vitale, Salvatore and Gerosa, Davide and Farr, Will M and Taylor, Stephen R},
  booktitle={Handbook of Gravitational Wave Astronomy},
  pages={1--60},
  year={2022},
  publisher={Springer}
}

@article{loredo2004accounting,
  author = {Loredo, Thomas J.},
  title = {Accounting for Source Uncertainties in Analyses of Astronomical Survey Data},
  journal = {AIP Conference Proceedings},
  volume = {735},
  pages = {195--206},
  year = {2004}
}

@article{messick2017analysis,
  title={Analysis framework for the prompt discovery of compact binary mergers in gravitational-wave data},
  author={Messick, Cody and Blackburn, Kent and Brady, Patrick and Brockill, Patrick and Cannon, Kipp and Cariou, Romain and Caudill, Sarah and Chamberlin, Sydney J and Creighton, Jolien DE and Everett, Ryan and others},
  journal={Physical Review D},
  volume={95},
  number={4},
  pages={042001},
  year={2017},
  publisher={APS}
}

@article{guo2020luminosity,
  title={The luminosity distribution of short gamma-ray bursts under a structured jet scenario},
  author={Guo, Qi and Wei, Daming and Wang, Yuanzhu},
  journal={The Astrophysical Journal},
  volume={894},
  number={1},
  pages={11},
  year={2020},
  publisher={The American Astronomical Society}
}

@article{von2020fourth,
  title={The fourth fermi-gbm gamma-ray burst catalog: A decade of data},
  author={Von Kienlin, Andreas and Meegan, Charles A and Paciesas, William S and Bhat, PN and Bissaldi, E and Briggs, MS and Burns, E and Cleveland, WH and Gibby, MH and Giles, MM and others},
  journal={The Astrophysical Journal},
  volume={893},
  number={1},
  pages={46},
  year={2020},
  publisher={The American Astronomical Society}
}

@article{gruber2014fermi,
  title={The Fermi GBM gamma-ray burst spectral catalog: four years of data},
  author={Gruber, David and Goldstein, Adam and von Ahlefeld, Victoria Weller and Bhat, P Narayana and Bissaldi, Elisabetta and Briggs, Michael S and Byrne, Dave and Cleveland, William H and Connaughton, Valerie and Diehl, Roland and others},
  journal={The Astrophysical Journal Supplement Series},
  volume={211},
  number={1},
  pages={12},
  year={2014},
  publisher={The American Astronomical Society}
}

@article{von2014second,
  title={The second Fermi GBM gamma-ray burst catalog: the first four years},
  author={Von Kienlin, Andreas and Meegan, Charles A and Paciesas, William S and Bhat, PN and Bissaldi, Elisabetta and Briggs, Michael S and Burgess, J Michael and Byrne, David and Chaplin, Vandiver and Cleveland, William and others},
  journal={The Astrophysical Journal Supplement Series},
  volume={211},
  number={1},
  pages={13},
  year={2014},
  publisher={The American Astronomical Society}
}

@article{bhat2016third,
  title={The third Fermi GBM gamma-ray burst catalog: the first six years},
  author={Bhat, P Narayana and Meegan, Charles A and Von Kienlin, Andreas and Paciesas, William S and Briggs, Michael S and Burgess, J Michael and Burns, Eric and Chaplin, Vandiver and Cleveland, William H and Collazzi, Andrew C and others},
  journal={The Astrophysical Journal Supplement Series},
  volume={223},
  number={2},
  pages={28},
  year={2016},
  publisher={The American Astronomical Society}
}

@article{salafia2023short,
  title={The short gamma-ray burst population in a quasi-universal jet scenario},
  author={Salafia, Om Sharan and Ravasio, Maria Edvige and Ghirlanda, Giancarlo and Mandel, Ilya},
  journal={Astronomy \& Astrophysics},
  volume={680},
  pages={A45},
  year={2023},
  publisher={EDP Sciences}
}

@article{wanderman2015rate,
  title={The rate, luminosity function and time delay of non-Collapsar short GRBs},
  author={Wanderman, David and Piran, Tsvi},
  journal={Monthly Notices of the Royal Astronomical Society},
  volume={448},
  number={4},
  pages={3026--3037},
  year={2015},
  publisher={Oxford University Press}
}

@article{tan2020jet,
  title={The jet structure and the intrinsic luminosity function of short gamma-ray bursts},
  author={Tan, Wei-Wei and Yu, Yun-Wei},
  journal={The Astrophysical Journal},
  volume={902},
  number={1},
  pages={83},
  year={2020},
  publisher={The American Astronomical Society}
}

@article{madau2014cosmic,
  title={Cosmic star-formation history},
  author={Madau, Piero and Dickinson, Mark},
  journal={Annual Review of Astronomy and Astrophysics},
  volume={52},
  number={1},
  pages={415--486},
  year={2014},
  publisher={Annual Reviews}
}

@article{vitale2019measuring,
  title={Measuring the star formation rate with gravitational waves from binary black holes},
  author={Vitale, Salvatore and Farr, Will M and Ng, Ken KY and Rodriguez, Carl L},
  journal={The Astrophysical Journal Letters},
  volume={886},
  number={1},
  pages={L1},
  year={2019},
  publisher={The American Astronomical Society}
}

@article{lan2023grb,
  title={GRB 221009a: an ordinary nearby GRB with extraordinary observational properties},
  author={Lan, Lin and Gao, He and Li, An and Xiao, Shuo and Ai, Shunke and Peng, Zong-Kai and Li, Long and Wang, Chen-Yu and Xu, Nan and Lin, Shijie and others},
  journal={The Astrophysical Journal Letters},
  volume={949},
  number={1},
  pages={L4},
  year={2023},
  publisher={The American Astronomical Society}
}

@article{barthelmy2005burst,
  title={The burst alert telescope (BAT) on the SWIFT midex mission},
  author={Barthelmy, Scott D and Barbier, Louis M and Cummings, Jay R and Fenimore, Ed E and Gehrels, Neil and Hullinger, Derek and Krimm, Hans A and Markwardt, Craig B and Palmer, David M and Parsons, Ann and others},
  journal={Space Science Reviews},
  volume={120},
  number={3},
  pages={143--164},
  year={2005},
  publisher={Springer}
}

@article{amati2002intrinsic,
  title={Intrinsic spectra and energetics of bepposax gamma--ray bursts with known redshifts},
  author={Amati, L and Frontera, Filippo and Tavani, M and Antonelli, A and Costa, E and Feroci, M and Guidorzi, C and Heise, J and Masetti, N and Montanari, E and others},
  journal={Astronomy \& Astrophysics},
  volume={390},
  number={1},
  pages={81--89},
  year={2002},
  publisher={EDP Sciences}
}

@article{yonetoku2004gamma,
  title={Gamma-ray burst formation rate inferred from the spectral peak energy-peak luminosity relation},
  author={Yonetoku, Daisuke and Murakami, Toshio and Nakamura, T and Yamazaki, Ryo and Inoue, AK and Ioka, K},
  journal={The Astrophysical Journal},
  volume={609},
  number={2},
  pages={935--951},
  year={2004}
}

@article{Karnesis:2023ras,
    author = "Karnesis, Nikolaos and Katz, Michael L. and Korsakova, Natalia and Gair, Jonathan R. and Stergioulas, Nikolaos",
    title = "{Eryn : A multi-purpose sampler for Bayesian inference}",
    eprint = "2303.02164",
    archivePrefix = "arXiv",
    primaryClass = "astro-ph.IM",
    month = "3",
    year = "2023"
}

@software{michael_katz_2023_7705496,
  author       = {Michael Katz and
                  Nikolaos Karnesis and
                  Natalia Korsakova},
  title        = {mikekatz04/Eryn: first full release},
  month        = mar,
  year         = 2023,
  publisher    = {Zenodo},
  version      = {v1.0.0},
  doi          = {10.5281/zenodo.7705496},
  url          = {https://doi.org/10.5281/zenodo.7705496}
}

@ARTICLE{2013PASP..125..306F,
       author = {{Foreman-Mackey}, Daniel and {Hogg}, David W. and {Lang}, Dustin and {Goodman}, Jonathan},
        title = "{emcee: The MCMC Hammer}",
      journal = {\pasp},
         year = 2013,
        month = mar,
       volume = {125},
       number = {925},
        pages = {306},
          doi = {10.1086/670067},
archivePrefix = {arXiv},
       eprint = {1202.3665},
 primaryClass = {astro-ph.IM},
       adsurl = {https://ui.adsabs.harvard.edu/abs/2013PASP..125..306F}
}

@software{jax2018github,
  author = {James Bradbury and Roy Frostig and Peter Hawkins and Matthew James Johnson and Chris Leary and Dougal Maclaurin and George Necula and Adam Paszke and Jake Vander{P}las and Skye Wanderman-{M}ilne and Qiao Zhang},
  title = {{JAX}: composable transformations of {P}ython+{N}um{P}y programs},
  url = {http://github.com/jax-ml/jax},
  version = {0.3.13},
  year = {2018},
}

@article{abac2025gwtc,
  title={GWTC-4.0: Population properties of merging compact binaries},
  author={Abac, AG and Abouelfettouh, I and Acernese, F and Ackley, K and Adamcewicz, C and Adhicary, S and Adhikari, D and Adhikari, N and Adhikari, RX and Adkins, VK and others},
  journal={arXiv preprint arXiv:2508.18083},
  year={2025}
}

@article{ligo2026gwtc,
  title={GWTC-5.0: Population Properties of Merging Compact Binaries},
  author={{The LIGO Scientific Collaboration} and {the Virgo Collaboration} and {the KAGRA Collaboration}},
  journal={arXiv preprint arXiv:2605.27226},
  year={2026}
}

@article{ray2023nonparametric,
  title={Nonparametric inference of the population of compact binaries from gravitational-wave observations using binned gaussian processes},
  author={Ray, Anarya and Hernandez, Ignacio Maga{\~n}a and Mohite, Siddharth and Creighton, Jolien and Kapadia, Shasvath},
  journal={The Astrophysical Journal},
  volume={957},
  number={1},
  pages={37},
  year={2023},
  publisher={The American Astronomical Society}
}

@article{yi2022gravitational,
  title={The Gravitational Wave Universe Toolbox-II. Constraining the binary black hole population with second and third generation detectors},
  author={Yi, Shu-Xu and Stoppa, Fiorenzo and Nelemans, Gijs and Cator, Eric},
  journal={Astronomy \& Astrophysics},
  volume={663},
  pages={A156},
  year={2022},
  publisher={EDP Sciences}
}

@article{yorgancioglu2025can,
  title={Can Gamma-Ray Burst Empirical Correlations Be Used for Population Studies?},
  author={Yorgancioglu, Emre S and Du, Yun-Fei and Yi, Shu-Xu and Zhang, Shuang-Nan},
  journal={The Astrophysical Journal},
  volume={989},
  number={2},
  pages={151},
  year={2025},
  publisher={The American Astronomical Society}
}

@article{yorgancioglu2025feasibility,
  title={On the feasibility of deriving pseudo-redshifts of gamma-ray bursts from two phenomenological correlations},
  author={Yorgancioglu, Emre S and Du, Yun-Fei and Yi, Shu-Xu and Moradi, Rahim and Feng, Hua and Zhang, Shuang-Nan},
  journal={The Astrophysical Journal},
  volume={981},
  number={2},
  pages={197},
  year={2025},
  publisher={The American Astronomical Society}
}

@article{collazzi2012significant,
  title={A significant problem with using the Amati relation for cosmological purposes},
  author={Collazzi, Andrew C and Schaefer, Bradley E and Goldstein, Adam and Preece, Robert D},
  journal={The Astrophysical Journal},
  volume={747},
  number={1},
  pages={39},
  year={2012},
  publisher={The American Astronomical Society}
}

@article{huang2021reconciling,
  title={Reconciling low and high redshift GRB luminosity correlations},
  author={Huang, Lu and Huang, Zhiqi and Luo, Xiaolin and He, Xinbo and Fang, Yuhong},
  journal={Physical Review D},
  volume={103},
  number={12},
  pages={123521},
  year={2021},
  publisher={APS}
}

@article{abbott2023population,
  title={Population of merging compact binaries inferred using gravitational waves through GWTC-3},
  author={Abbott, Richard and Abbott, TD and Acernese, F and Ackley, K and Adams, C and Adhikari, N and Adhikari, RX and Adya, VB and Affeldt, C and Agarwal, D and others},
  journal={Physical Review X},
  volume={13},
  number={1},
  pages={011048},
  year={2023},
  publisher={APS}
}

@article{berger2014short,
  title={Short-duration gamma-ray bursts},
  author={Berger, Edo},
  journal={Annual review of Astronomy and Astrophysics},
  volume={52},
  number={1},
  pages={43--105},
  year={2014},
  publisher={Annual Reviews}
}

@article{abbott2017gw170817,
  title={GW170817: observation of gravitational waves from a binary neutron star inspiral},
  author={Abbott, Benjamin P and Abbott, Rich and Abbott, Thomas D and Acernese, Fausto and Ackley, Kendall and Adams, Carl and Adams, Thomas and Addesso, Paolo and Adhikari, Rana X and Adya, Vaishali B and others},
  journal={Physical review letters},
  volume={119},
  number={16},
  pages={161101},
  year={2017},
  publisher={APS}
}

@article{ghirlanda2016short,
  title={Short gamma-ray bursts at the dawn of the gravitational wave era},
  author={Ghirlanda, Giancarlo and Salafia, Om Sharan and Pescalli, A and Ghisellini, Gabriele and Salvaterra, Ruben and Chassande--Mottin, E and Colpi, M and Nappo, F and D’Avanzo, Paolo and Melandri, A and others},
  journal={Astronomy \& Astrophysics},
  volume={594},
  pages={A84},
  year={2016},
  publisher={EDP Sciences}
}

@article{drout2017light,
  title={Light curves of the neutron star merger GW170817/SSS17a: Implications for r-process nucleosynthesis},
  author={Drout, Maria R and Piro, AL and Shappee, BJ and Kilpatrick, CD and Simon, JD and Contreras, C and Coulter, DA and Foley, RJ and Siebert, MR and Morrell, N and others},
  journal={Science},
  volume={358},
  number={6370},
  pages={1570--1574},
  year={2017},
  publisher={American Association for the Advancement of Science}
}

@article{radice2020dynamics,
  title={The dynamics of binary neutron star mergers and GW170817},
  author={Radice, David and Bernuzzi, Sebastiano and Perego, Albino},
  journal={Annual Review of Nuclear and Particle Science},
  volume={70},
  number={1},
  pages={95--119},
  year={2020},
  publisher={Annual Reviews}
}

@article{fong2015decade,
  title={A decade of short-duration gamma-ray burst broadband afterglows: energetics, circumburst densities, and jet opening angles},
  author={Fong, Wen-fai and Berger, Edo and Margutti, Raffaella and Zauderer, B Ashley},
  journal={The Astrophysical Journal},
  volume={815},
  number={2},
  pages={102},
  year={2015},
  publisher={The American Astronomical Society}
}

@article{rouco2023jet,
  title={The jet opening angle and event rate distributions of short gamma-ray bursts from late-time x-ray afterglows},
  author={Rouco Escorial, Alicia and Fong, Wen-fai and Berger, Edo and Laskar, Tanmoy and Margutti, Raffaella and Schroeder, Genevieve and Rastinejad, Jillian C and Cornish, Dylaan and Popp, Sarah and Lally, Maura and others},
  journal={The Astrophysical Journal},
  volume={959},
  number={1},
  pages={13},
  year={2023},
  publisher={The American Astronomical Society}
}

@article{mancarella2025sampling,
  title={Sampling the full hierarchical population posterior distribution in gravitational-wave astronomy},
  author={Mancarella, Michele and Gerosa, Davide},
  journal={Physical Review D},
  volume={111},
  number={10},
  pages={103012},
  year={2025},
  publisher={APS}
}

@article{wysocki2019reconstructing,
  title={Reconstructing phenomenological distributions of compact binaries via gravitational wave observations},
  author={Wysocki, Daniel and Lange, Jacob and O’Shaughnessy, Richard},
  journal={Physical Review D},
  volume={100},
  number={4},
  pages={043012},
  year={2019},
  publisher={APS}
}

@article{doctor2020black,
  title={Black hole coagulation: Modeling hierarchical mergers in black hole populations},
  author={Doctor, Zoheyr and Wysocki, Daniel and O’Shaughnessy, Richard and Holz, Daniel E and Farr, Ben},
  journal={The Astrophysical Journal},
  volume={893},
  number={1},
  pages={35},
  year={2020},
  publisher={The American Astronomical Society}
}

@article{delfavero2021normal,
  title={Normal Approximate Likelihoods to Gravitational Wave Events},
  author={Delfavero, Vera and O'Shaughnessy, Richard and Wysocki, Daniel and Yelikar, Anjali},
  journal={arXiv preprint arXiv:2107.13082},
  year={2021}
}

@article{golomb2022hierarchical,
  title={Hierarchical inference of binary neutron star mass distribution and equation of state with gravitational waves},
  author={Golomb, Jacob and Talbot, Colm},
  journal={The Astrophysical Journal},
  volume={926},
  number={1},
  pages={79},
  year={2022},
  publisher={The American Astronomical Society}
}

@article{mould2024calibrating,
  title={Calibrating signal-to-noise ratio detection thresholds using gravitational-wave catalogs},
  author={Mould, Matthew and Moore, Christopher J and Gerosa, Davide},
  journal={Physical Review D},
  volume={109},
  number={6},
  pages={063013},
  year={2024},
  publisher={APS}
}

@article{hussain2026hints,
  title={Hints of spin-magnitude correlations and a rapidly spinning subpopulation of binary black holes},
  author={Hussain, Asad and Isi, Maximiliano and Zimmerman, Aaron},
  journal={The Astrophysical Journal},
  volume={996},
  number={1},
  pages={71},
  year={2026},
  publisher={The American Astronomical Society}
}

@article{farr2019accuracy,
  title={Accuracy requirements for empirically-measured selection functions},
  author={Farr, Will M},
  journal={arXiv preprint arXiv:1904.10879},
  year={2019}
}

@article{lattimer1974black,
  title={Black-hole-neutron-star collisions},
  author={Lattimer, James M and Schramm, David N},
  journal={Astrophysical Journal, vol. 192, Sept. 15, 1974, pt. 2, p. L145-L147.},
  volume={192},
  pages={L145--L147},
  year={1974}
}

@article{eichler1989nucleosynthesis,
  title={Nucleosynthesis, neutrino bursts and $\gamma$-rays from coalescing neutron stars},
  author={Eichler, David and Livio, Mario and Piran, Tsvi and Schramm, David N},
  journal={Nature},
  volume={340},
  number={6229},
  pages={126--128},
  year={1989},
  publisher={Nature Publishing Group UK London}
}

@article{rosswog1998mass,
  title={Mass ejection in neutron star mergers},
  author={Rosswog, Stephan and Liebend{\"o}rfer, M and Thielemann, F-K and Davies, MB and Benz, W and Piran, T},
  journal={Arxiv preprint astro-ph/9811367},
  year={1998}
}

@article{dominik2012double,
  title={Double compact objects. I. The significance of the common envelope on merger rates},
  author={Dominik, Michal and Belczynski, Krzysztof and Fryer, Christopher and Holz, Daniel E and Berti, Emanuele and Bulik, Tomasz and Mandel, Ilya and O'shaughnessy, Richard},
  journal={The Astrophysical Journal},
  volume={759},
  number={1},
  pages={52},
  year={2012},
  publisher={The American Astronomical Society}
}

@article{dominik2013double,
  title={Double compact objects. II. Cosmological merger rates},
  author={Dominik, Michal and Belczynski, Krzysztof and Fryer, Christopher and Holz, Daniel E and Berti, Emanuele and Bulik, Tomasz and Mandel, Ilya and O'Shaughnessy, Richard},
  journal={The Astrophysical Journal},
  volume={779},
  number={1},
  pages={72},
  year={2013},
  publisher={The American Astronomical Society}
}

@article{belczynski2016first,
  title={The first gravitational-wave source from the isolated evolution of two stars in the 40--100 solar mass range},
  author={Belczynski, Krzysztof and Holz, Daniel E and Bulik, Tomasz and O’Shaughnessy, Richard},
  journal={Nature},
  volume={534},
  number={7608},
  pages={512--515},
  year={2016},
  publisher={Nature Publishing Group UK London}
}

@article{nelemans2001gravitational,
  title={The gravitational wave signal from the Galactic disk population of binaries containing two compact objects},
  author={Nelemans, Gijs and Yungelson, LR and Zwart, SF Portegies},
  journal={Astronomy \& Astrophysics},
  volume={375},
  number={3},
  pages={890--898},
  year={2001},
  publisher={EDP Sciences}
}

@article{nelemans2009galactic,
  title={The Galactic gravitational wave foreground},
  author={Nelemans, Gijs},
  journal={Classical and Quantum Gravity},
  volume={26},
  number={9},
  pages={094030},
  year={2009}
}

@article{abadie2010predictions,
  title={Predictions for the rates of compact binary coalescences observable by ground-based gravitational-wave detectors},
  author={Abadie, Jea and Abbott, BP and Abbott, Richard and Abernathy, Matthew and Accadia, Timothee and Acernese, Fausto and Adams, Carl and Adhikari, Rana and Ajith, Parameswaran and Allen, Bruce and others},
  journal={Classical and Quantum Gravity},
  volume={27},
  number={17},
  pages={173001},
  year={2010}
}

@article{morinaga2019statistical,
  title={Statistical properties of substructures around Milky Way-sized haloes and their implications for the formation of stellar streams},
  author={Morinaga, Yu and Ishiyama, Tomoaki and Kirihara, Takanobu and Kinjo, Kazuki},
  journal={Monthly Notices of the Royal Astronomical Society},
  volume={487},
  number={2},
  pages={2718--2729},
  year={2019},
  publisher={Oxford University Press}
}

@article{abbott2021population,
  title={Population properties of compact objects from the second LIGO--Virgo gravitational-wave transient catalog},
  author={Abbott, Rich and Abbott, TD and Abraham, S and Acernese, Fausto and Ackley, K and Adams, A and Adams, C and Adhikari, RX and Adya, VB and Affeldt, Christoph and others},
  journal={The Astrophysical journal letters},
  volume={913},
  number={1},
  pages={L7},
  year={2021},
  publisher={The American Astronomical Society}
}

@article{jakobsson2006mean,
  title={A mean redshift of 2.8 for Swift gamma-ray bursts},
  author={Jakobsson, P and Levan, A and Fynbo, JPU and Priddey, R and Hjorth, J and Tanvir, N and Watson, D and Jensen, BL and Sollerman, J and Natarajan, P and others},
  journal={Astronomy \& Astrophysics},
  volume={447},
  number={3},
  pages={897--903},
  year={2006},
  publisher={EDP sciences}
}

@article{fynbo2009low,
  title={Low-resolution spectroscopy of gamma-ray burst optical afterglows: biases in the Swift sample and characterization of the absorbers},
  author={Fynbo, Johan Peter Uldall and Jakobsson, Pall and Prochaska, Jason Xavier and Malesani, D and Ledoux, Cedric and de Ugarte Postigo, Antonio and Nardini, Marco and Vreeswijk, Paul Marijn and Wiersema, Klass and Hjorth, Jens and others},
  journal={The Astrophysical Journal Supplement Series},
  volume={185},
  number={2},
  pages={526--573},
  year={2009},
  publisher={The American Astronomical Society}
}

@article{d2015short,
  title={Short gamma-ray bursts: A review},
  author={D'Avanzo, Paolo},
  journal={Journal of High Energy Astrophysics},
  volume={7},
  pages={73--80},
  year={2015},
  publisher={Elsevier}
}

@article{rastinejad2022kilonova,
  title={A kilonova following a long-duration gamma-ray burst at 350 Mpc},
  author={Rastinejad, Jillian C and Gompertz, Benjamin P and Levan, Andrew J and Fong, Wen-fai and Nicholl, Matt and Lamb, Gavin P and Malesani, Daniele B and Nugent, Anya E and Oates, Samantha R and Tanvir, Nial R and others},
  journal={Nature},
  volume={612},
  number={7939},
  pages={223--227},
  year={2022},
  publisher={Nature Publishing Group UK London}
}

@article{levan2024heavy,
  title={Heavy-element production in a compact object merger observed by JWST},
  author={Levan, Andrew J and Gompertz, Benjamin P and Salafia, Om Sharan and Bulla, Mattia and Burns, Eric and Hotokezaka, Kenta and Izzo, Luca and Lamb, Gavin P and Malesani, Daniele B and Oates, Samantha R and others},
  journal={Nature},
  volume={626},
  number={8000},
  pages={737--741},
  year={2024},
  publisher={Nature Publishing Group UK London}
}

@article{yang2022boosting,
  title={Boosting K-ion kinetics by interfacial polarization induced by amorphous MoO3-x for MoSe2/MoO3-x@ rGO composites},
  author={Yang, Jiangshao and Liu, Liwen and Wang, Daoyi and Tao, Jianming and Yang, Yanming and Li, Jiaxin and Lin, Yingbin and Huang, Zhigao},
  journal={Journal of Materials Science \& Technology},
  volume={115},
  pages={232--240},
  year={2022},
  publisher={Elsevier}
}

@article{sun2025magnetar,
  title={Magnetar emergence in a peculiar gamma-ray burst from a compact star merger},
  author={Sun, Hui and Wang, Chenwei and Yang, Jun and Zhang, Bin-Bin and Xiong, Shaolin and Yin, Yi-Han Iris and Liu, Yuan and Li, Ye and Xue, Wangchen and Yan, Zhen-Yu and others},
  journal={National Science Review},
  volume={12},
  number={3},
  pages={nwae401},
  year={2025},
  publisher={Oxford University Press}
}

@article{wang2025subclass,
  title={A Subclass of Gamma-Ray Burst Originating from Compact Binary Merger},
  author={Wang, Chen-Wei and Tan, Wen-Jun and Xiong, Shao-Lin and Yi, Shu-Xu and Moradi, Rahim and Li, Bing and Zhang, Zhen and Wang, Yu and Meng, Yan-Zhi and Wu, Bo-Bing and others},
  journal={The Astrophysical Journal},
  volume={979},
  number={1},
  pages={73},
  year={2025},
  publisher={The American Astronomical Society}
}

@article{yi2025long,
  title={Long pulse by short central engine: Prompt emission from expanding dissipation rings in the jet front of gamma-ray bursts},
  author={Yi, Shu-Xu and Yorgancioglu, Emre Seyit and Xiong, S-L and Zhang, S-N},
  journal={Journal of High Energy Astrophysics},
  volume={47},
  pages={100359},
  year={2025},
  publisher={Elsevier}
}

@article{narayan1992gamma,
  title={Gamma-ray bursts as the death throes of massive binary stars},
  author={Narayan, Ramesh and Paczy{\'n}ski, Bohdan and Piran, Tsvi},
  journal={arXiv preprint astro-ph/9204001},
  year={1992}
}

@article{nakar2007short,
  title={Short-hard gamma-ray bursts},
  author={Nakar, Ehud},
  journal={Physics Reports},
  volume={442},
  number={1-6},
  pages={166--236},
  year={2007},
  publisher={Elsevier}
}

@article{hurley2005exceptionally,
  title={An exceptionally bright flare from SGR 1806--20 and the origins of short-duration $\gamma$-ray bursts},
  author={Hurley, K and Boggs, SE and Smith, DM and Duncan, RC and Lin, R and Zoglauer, A and Krucker, S and Hurford, G and Hudson, H and Wigger, C and others},
  journal={Nature},
  volume={434},
  number={7037},
  pages={1098--1103},
  year={2005},
  publisher={Nature Publishing Group UK London}
}

@article{bromberg2013short,
  title={Short versus long and collapsars versus non-collapsars: a quantitative classification of gamma-ray bursts},
  author={Bromberg, Omer and Nakar, Ehud and Piran, Tsvi and Sari, Re'em},
  journal={The Astrophysical Journal},
  volume={764},
  number={2},
  pages={179},
  year={2013},
  publisher={The American Astronomical Society}
}

@article{mooley2018superluminal,
  title={Superluminal motion of a relativistic jet in the neutron-star merger GW170817},
  author={Mooley, KP and Deller, AT and Gottlieb, O and Nakar, E and Hallinan, G and Bourke, S and Frail, DA and Horesh, A and Corsi, A and Hotokezaka, K},
  journal={Nature},
  volume={561},
  number={7723},
  pages={355--359},
  year={2018},
  publisher={Nature Publishing Group UK London}
}

@ARTICLE{2004ApJ...601L.179K,
       author = {{Kalogera}, V. and {Kim}, C. and {Lorimer}, D.~R. and {Burgay}, M. and {D'Amico}, N. and {Possenti}, A. and {Manchester}, R.~N. and {Lyne}, A.~G. and {Joshi}, B.~C. and {McLaughlin}, M.~A. and {Kramer}, M. and {Sarkissian}, J.~M. and {Camilo}, F.},
        title = "{The Cosmic Coalescence Rates for Double Neutron Star Binaries}",
      journal = {\apjl},
         year = 2004,
        month = feb,
       volume = {601},
       number = {2},
        pages = {L179-L182},
          doi = {10.1086/382155},
archivePrefix = {arXiv},
       eprint = {astro-ph/0312101},
 primaryClass = {astro-ph},
       adsurl = {https://ui.adsabs.harvard.edu/abs/2004ApJ...601L.179K}
}

@ARTICLE{2019ApJ...880L...8A,
       author = {{Andrews}, Jeff J. and {Mandel}, Ilya},
        title = "{Double Neutron Star Populations and Formation Channels}",
      journal = {\apjl},
         year = 2019,
        month = jul,
       volume = {880},
       number = {1},
          eid = {L8},
        pages = {L8},
          doi = {10.3847/2041-8213/ab2ed1},
archivePrefix = {arXiv},
       eprint = {1904.12745},
 primaryClass = {astro-ph.HE},
       adsurl = {https://ui.adsabs.harvard.edu/abs/2019ApJ...880L...8A}
}

@ARTICLE{2019MNRAS.486.2896S,
       author = {{Simonetti}, Paolo and {Matteucci}, Francesca and {Greggio}, Laura and {Cescutti}, Gabriele},
        title = "{A new delay time distribution for merging neutron stars tested against Galactic and cosmic data}",
      journal = {\mnras},
         year = 2019,
        month = jun,
       volume = {486},
       number = {2},
        pages = {2896-2909},
          doi = {10.1093/mnras/stz991},
archivePrefix = {arXiv},
       eprint = {1901.02016},
 primaryClass = {astro-ph.GA},
       adsurl = {https://ui.adsabs.harvard.edu/abs/2019MNRAS.486.2896S}
}

@ARTICLE{2019MNRAS.487.4847B,
       author = {{Beniamini}, Paz and {Piran}, Tsvi},
        title = "{The gravitational waves merger time distribution of binary neutron star systems}",
      journal = {\mnras},
         year = 2019,
        month = aug,
       volume = {487},
       number = {4},
        pages = {4847-4854},
          doi = {10.1093/mnras/stz1589},
archivePrefix = {arXiv},
       eprint = {1903.11614},
 primaryClass = {astro-ph.HE},
       adsurl = {https://ui.adsabs.harvard.edu/abs/2019MNRAS.487.4847B}
}

@article{coward2012swift,
  title={The Swift short gamma-ray burst rate density: implications for binary neutron star merger rates},
  author={Coward, DM and Howell, EJ and Piran, Tsvi and Stratta, Giulia and Branchesi, Marica and Bromberg, Omer and Gendre, Bruce and Burman, RR and Guetta, Dafne},
  journal={Monthly Notices of the Royal Astronomical Society},
  volume={425},
  number={4},
  pages={2668--2673},
  year={2012},
  publisher={Blackwell Science Ltd Oxford, UK}
}

@article{goldstein2012fermi,
  title={The Fermi GBM gamma-ray burst spectral catalog: the first two years},
  author={Goldstein, Adam and Burgess, J Michael and Preece, Robert D and Briggs, Michael S and Guiriec, Sylvain and van der Horst, Alexander J and Connaughton, Valerie and Wilson-Hodge, Colleen A and Paciesas, William S and Meegan, Charles A and others},
  journal={The Astrophysical Journal Supplement Series},
  volume={199},
  number={1},
  pages={19},
  year={2012},
  publisher={The American Astronomical Society}
}

@ARTICLE{2026ApJ..1006L..49F,
       author = {{Fishbach}, Maya and {Ji}, Alexander P. and {Fong}, Wen-fai and {Wu}, Tom Y. and {Rastinejad}, Jillian C. and {Vijaykumar}, Aditya and {Chen}, Hsin-Yu},
        title = "{Implications of Low Neutron Star Merger Rates for Gamma-Ray Bursts, r-process Production, and Galactic Double Neutron Stars}",
      journal = {\apjl},
         year = 2026,
        month = aug,
       volume = {1006},
       number = {2},
          eid = {L49},
        pages = {L49},
          doi = {10.3847/2041-8213/ae8756},
archivePrefix = {arXiv},
       eprint = {2604.05059},
 primaryClass = {astro-ph.HE},
       adsurl = {https://ui.adsabs.harvard.edu/abs/2026ApJ..1006L..49F}
}
\bibliographystyle{aasjournalv7}



\end{document}